# Interpretable data-driven turbulence modeling for separated flows using symbolic regression with unit constraints

Boqian Zhang[1] and Juanmian Lei[1, a)]

[1] School of Aerospace Engineering, Beijing Institute of Technology, Beijing, 100081, China

Machine learning techniques have been applied to enhance turbulence modeling in recent years. However, the "black box" nature of most machine learning techniques poses significant interpretability challenges in improving turbulence models. This paper introduces a novel unit-constrained turbulence modeling framework using symbolic regression to overcome these challenges. The framework amends the constitutive equation of linear eddy viscosity models (LEVMs) by establishing explicit equations between the Reynolds stress deviation and mean flow quantities, thereby improving the LEVM model's predictive capability for large separated turbulence. Unit consistency constraints are applied to the symbolic expressions to ensure physical realizability. The effectiveness of the framework and the generalization capability of the learned model are demonstrated through its application to the separated flow over 2D periodic hills and a backward-facing step. Compared to the standard *k*-*ε* model, the learned model shows significantly improved predictive accuracy for anisotropic Reynolds stresses, velocity and skin friction, while exhibiting promising generalization capabilities across various scenarios.

**Keywords:** Turbulence modeling; Symbolic regression; Separated flow; Machine learning

## 1. Introduction

Turbulence is prevalent across natural phenomena and engineering applications, and Computational Fluid Dynamics (CFD) is an essential tool for studying turbulence. Achieving accurate and efficient numerical simulations of turbulent flows is a significant issue that persists in both academic and industrial spheres. Over the past few decades, a variety of numerical simulation techniques for turbulence have been developed, including Direct Numerical Simulation (DNS), Large Eddy Simulation (LES), and Reynolds-Averaged Navier-Stokes (RANS) approaches, with RANS being the most extensively researched and applied. Despite earlier predictions suggesting that LES might supplant RANS in industrial applications within the forthcoming decades [1], such optimistic expectations have been

a) Corresponding author, E-mail: leijm@bit.edu.cn

shattered with the end of "Moore's Law". At least in the near future, RANS will continue to be the primary tool for industrial CFD applications [1, 2].

The RANS method employs the Reynolds-averaged Navier–Stokes equations as the governing equations. A significant challenge within these equations is the non-closed Reynolds stress term, necessitating the development of constitutive equations to elucidate the relationship between the Reynolds stress and the time-averaged flow field variables, thereby achieving closure of the equations. This requirement has led to the creation of various turbulence closure models. Among these, linear eddy viscosity models (LEVMs), based on the Boussinesq assumption, are distinguished by their low computational demands and robust performance, making them a popular choice in engineering applications [3–5]. The Boussinesq assumption posits a linear relationship between the anisotropic Reynolds stress tensor and the time-averaged flow strain rate tensor. However, this assumption has been proven to diverge from reality in most real-world flows [6], especially in flows with large separation, secondary flows, and curvature, etc., thereby undermining the accuracy of LEVMs in simulating such flows [7]. In contrast to LEVMs, the Reynolds stress model (RSM) and the algebraic stress model (ASM), grounded in the Reynolds stress transport equation, do not rely on the Boussinesq assumption and offer enhanced predictive accuracy for complex turbulence. However, their application has been limited by numerical instability and elevated computational costs [8].

With the advancements in machine learning technologies and the increasing availability of high-precision turbulence data, data-driven turbulence modeling that integrates data with physical prior has emerged as a novel paradigm in turbulence modeling. This paradigm has been successfully applied to enhance LEVMs. Duraisamy et al. [9, 10] proposed a turbulence modeling method based on field inversion and machine learning (FIML). They first obtain the spatial distribution of model correction function through field inversion, then use machine learning to establish the mapping relationship between flow variables and the correction parameter, thereby enhancing the turbulence model's predictive capabilities. Ling et al. [11] developed a tensor basis neural network (TBNN) incorporating Galilean invariance to predict the anisotropic Reynolds stress tensor accurately. Wang et al. [2] established the mapping relationship between the mean flow variables and the Reynolds stress tensor based on random forest method, improving the original turbulence model's prediction accuracy for fully developed turbulence in square pipe and large separated flows. Wu et al. [12] further established a comprehensive physics-informed machine learning framework for turbulence modeling, addressing the ill-conditioned problem of RANS equations. Shan et al. [13]

employed a deep neural network to predict the eddy viscosity within the S-A turbulence model, reducing reliance on high-precision turbulence experimental data during model training using data assimilation technology.

Traditionally, research in data-driven turbulence modeling usually establishes the mapping relationship between the average flow variables and Reynolds stress or other turbulence quantities through techniques such as neural networks, random forests, etc. However, these mapping relationships tend to be unobservable and uninterpretable, presenting a significant challenge to the interpretability of machine learning applications in turbulence modeling. This lack of interpretability is a critical barrier that requires resolution for the effective application of machine learning in turbulence modeling [14]. In recent years, a unique technique—symbolic regression—has attracted considerable attention. Unlike other "black-box" models, symbolic regression aims to derive explicit equations that link input variables $x_i$ with a target variable $y$. The explicit nature of these equations allows for a straightforward interpretation of their physical meaning, and this transparency facilitates the integration of physical prior knowledge, thereby enhancing both the interpretability and extrapolation capabilities of the models [12].

Symbolic regression has been applied to address classical closure problems in fluid mechanics, including modeling of turbulence [15–22], constitutive for non-Newtonian fluids [23], multi-physics coupling [24], and fluid transport properties [25]. In turbulence modeling, symbolic regression offers an additional advantage: its explicit equations can be directly integrated into CFD solvers, thereby mitigating the ill-conditioning issues that often affect traditional data-driven models [8, 12]. This underscores the tremendous potential of symbolic regression in turbulence modeling. Weatheritt and Sandberg [15] employed a genetic programming (GP) based method to derive an explicit algebraic expression for the Reynolds stresses, successfully mitigating the SST model's tendency to overpredict flow separation. Wu et al. [16, 17] further improved the separation flow predictions of the SST model by combining evolutionary algorithms with flow field inversion techniques, and Tang et al. [18] later extended this approach to hypersonic flows. Schmelzer et al. [19] introduced the SpaRTA method based on sparse regression, which enhanced the SST model's accuracy by leveraging sparsity promotion techniques; subsequently, Ben et al. [20] and Stöcker et al. [21] further advanced this method. More recently, deep learning–based symbolic regression methods have also been applied in Reynolds stress modeling [22], demonstrating promising results.

Despite the significant promise of symbolic regression in turbulence modeling, traditional algorithms in this domain, particularly those based on GP, encounter challenges related to extensive search spaces and slow learning speeds [26]. These limitations constrict the size of training datasets and the number of input features, consequently limiting the complexity of the symbolic expressions that can be generated. This, in turn, limits the predictive capabilities of the learned turbulence models, and poses obstacles for more refined modeling research. Research shows that within the search space of conventional symbolic regression algorithms, most candidate expressions fail to adhere to the principle of unit consistency—that is, the units across equations or within addition and subtraction operations do not align [27]. Such inconsistencies render these expressions physically meaningless, leading to considerable inefficiencies in model learning. However, in turbulence modeling, it is a common practice to normalize flow quantities by the turbulence scale for modeling purposes. However, this normalization process results in the omission of unit information, which could otherwise be beneficial in data-driven turbulence modeling. The absence of these constraints leads to an expanded search space for algorithms, which adversely affects the learning efficiency and the interpretability of the models developed.

To address the limitations above, a unit-constrained turbulence modeling framework using symbolic regression is proposed. Within this framework, a new learned turbulence model for separated flow is developed. The performance and generalization capabilities of the learned model are then evaluated. The structure of this paper is as follows: Section 2 outlines the turbulence closure problem, and introduces the proposed unit-constrained turbulence modeling framework using symbolic regression and the numerical methods used in this study. Section 3 details the training case, input features, hyperparameters, and mesh configurations. Section 4 applies the proposed turbulence modeling framework to the cases of flows over periodic hills and a backward-facing step. Section 5 presents the conclusions and outlook on future directions.

## 2. Methodology

In this section, we first introduce the turbulence closure issue and baseline turbulence model. Then, we introduce the proposed unit-constrained turbulence modeling framework using symbolic regression, detailing the algorithm structure and modeling process, and introduce the implementation method of unit constraint. Finally, the CFD solver and numerical methods used in the numerical simulations are described.

### 2.1. Turbulence closure

The Reynolds-averaged Navier-Stokes equation is the governing equation for RANS simulations. For incompressible flow, the equation is given by:

$$\frac{\partial \mathbf{U}}{\partial t}+\mathbf{U}\cdot\nabla\mathbf{U}=-\frac{1}{\rho}\nabla p+\nu\nabla^2\mathbf{U}+\nabla\cdot\boldsymbol{\tau} \tag{1}$$

where $\boldsymbol{U}$ is the time-averaged velocity vector, $\rho$ represents density, $p$ is the time-averaged pressure, $\nu$ is the kinematic viscosity, and $\boldsymbol{\tau}$ is the Reynolds stress tensor. The Reynolds stress $\boldsymbol{\tau}$ is a 3×3 symmetric tensor. For statistically two-dimensional flows, the Reynolds stress has four independent and non-zero components, which are $\tau_{11}$, $\tau_{22}$, $\tau_{33}$ and $\tau_{12}\left(=\tau_{21}\right)$, where $\tau_{33}$ is non-zero due to turbulent spanwise fluctuations. For three-dimensional flows, the Reynolds stress tensor has six independent and non-zero components, which are $\tau_{11}$ 、 $\tau_{22}$ 、 $\tau_{33}$ 、 $\tau_{12}\left(=\tau_{21}\right)$ 、 $\tau_{13}\left(=\tau_{31}\right)$ and $\tau_{23}(=\tau_{32})$.

To close Eq. 1, LEVMs based on the Boussinesq assumption assume that the anisotropic Reynolds stress tensor $\boldsymbol{a}$ is linearly related to the mean strain rate tensor $\boldsymbol{S}$, given by:

$$\mathbf{a}=\boldsymbol{\tau}-\frac{2}{3}k\delta_{ij}=-2\nu_t\mathbf{S} \tag{2}$$

where $k$ represents turbulent kinetic energy, $\delta_{ij}$ is the Kronecker delta, and $\nu_t$ is the turbulent viscosity. The value of $\nu_t$ is determined by the scale of turbulence and is calculated differently across various turbulence models.

Widely used LEVMs include the S-A turbulence model [28], the $k$-$\varepsilon$ model [29], the $k$-$\omega$ model [30], and the Shear Stress Transport (SST) model [31]. The standard $k$-$\varepsilon$ model, proposed by B.E. Launder and D.B. Spalding, is one of the most commonly used turbulence models in engineering and serves as a benchmark in many data-driven turbulence modeling studies [2]. In the standard $k$-$\varepsilon$ model, the value of $\nu_t$ is calculated using the following equation [7]:

$$\nu_t=C_\mu\frac{k^2}{\varepsilon} \tag{3}$$

where $\varepsilon$ is the turbulent dissipation rate, $C_\mu$ is an empirical coefficient typically set to 0.09. The transport equations for turbulent kinetic energy $k$ and turbulent dissipation rate ε are given by:

$$\frac{\partial k}{\partial t}+U_j\frac{\partial k}{\partial x_j}=\tau_{ij}\frac{\partial U_i}{\partial x_j}-\varepsilon+\frac{\partial}{\partial x_j}\left[\left(\nu+\frac{\nu_t}{\sigma_k}\right)\frac{\partial k}{\partial x_j}\right] \tag{4}$$

$$\frac{\partial \varepsilon}{\partial t}+U_j\frac{\partial \varepsilon}{\partial x_j}=C_{\varepsilon_1}\frac{\varepsilon}{k}\tau_{ij}\frac{\partial U_i}{\partial x_j}-C_{\varepsilon_2}\frac{\varepsilon^2}{k}+\frac{\partial}{\partial x_j}\left[\left(\nu+\frac{\nu_t}{\sigma_\varepsilon}\right)\frac{\partial \varepsilon}{\partial x_j}\right] \tag{5}$$

where $\mu$ is the dynamic viscosity, $\sigma_k$ 、 $C_{\varepsilon_1}$ 、 $C_{\varepsilon_2}$ and $\sigma_\varepsilon$ are model coefficients, typically set to:

$$\sigma_k=1.0,\ C_{\varepsilon_1}=1.44,\ C_{\varepsilon_2}=1.92,\ \sigma_\varepsilon=1.3 \tag{6}$$

In this study, the standard $k$-$\varepsilon$ model is used as the baseline turbulence model to compute the baseline flow field and serves as the foundation for symbolic regression turbulence modeling.

## 2.2. Unit-constrained turbulence modeling framework using symbolic regression

**1. Symbolic regression guided by unit constraint**

Symbolic regression is traditionally implemented through GP techniques [32]. In recent years, deep learning has increasingly been applied to solve symbolic regression problems [27, 33, 34]. Tenachi et al. [27] introduced the Φ-SO symbolic regression algorithm guided by unit constraints based on recurrent neural networks (RNNs). This algorithm demonstrates notable efficiency and accuracy in expression learning. The turbulence modeling framework proposed in this study is based on the Φ-SO symbolic regression algorithm.

The Φ-SO algorithm operates based on RNNs and requires inputs to be in sequence form. Therefore, the first step is to transform the expression to be predicted into a symbolic sequence. This transformation is facilitated through the construction of a symbolic tree. Figure 1 demonstrates this process using the basic formula for hydrostatic pressure as an example. As shown in Fig. 1a, the right side of the equation consists of four independent variables: atmospheric static pressure $p_a$, fluid density $\rho$, gravitational acceleration $g$, height $h$, and two operators $\{+, \times\}$. Those quantities and operators are collectively referred to as the tokens constituting the expression. The collection of all possible tokens that might contribute to the expression's formation is referred to as the tokens library. These tokens are organized into a tree structure based on their operational relationships, known as a symbolic tree, as shown in Fig. 1b. Each token within the symbolic tree is sequentially numbered from top to bottom and from left to right. By sequencing the tokens according to their assigned numbers, the corresponding symbolic sequence for the expression is derived, as illustrated in Fig. 1c. This process enables the representation of any expression as a corresponding symbolic sequence, thus rendering it compatible with processing by RNNs.

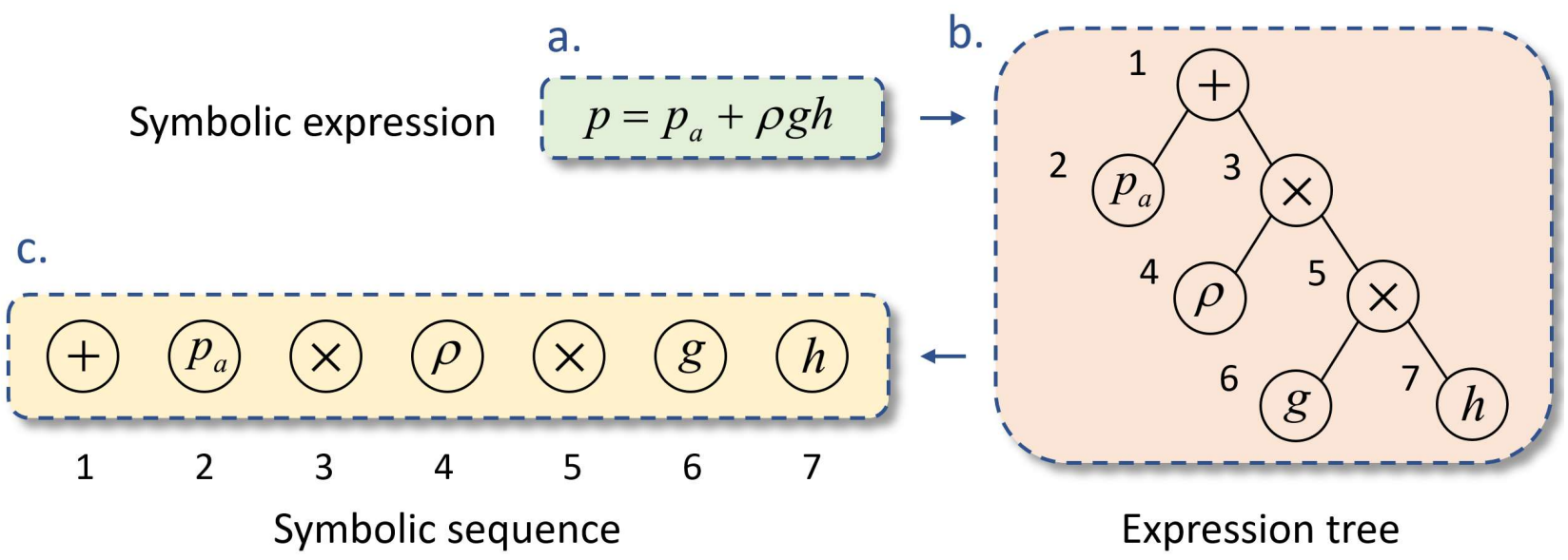

**Fig. 1** The process of converting a symbolic expression into a symbolic sequence through an expression tree

The architecture of the Φ-SO algorithm is shown in Fig. 2. Within the algorithm, each RNN cell is aligned with a specific position in the symbolic sequence, with the total number of RNN cells matching the maximum allowed length for an expression. Each RNN cell receives an observation as input for its position, which includes the parent token and its unit, sibling token and its unit, previous token and its unit, the unit required at the current position, and the minimum number of operators needed to complete the expression. Taking the expression shown in Fig. 1 as an example, the third position in Fig. 1a corresponds to the third RNN cell (counting from left to right). It can be seen from Fig. 1b, the input (observation) for this RNN cell is the parent token "+", the sibling token " $p_a$", and the previous token "$p_a$", with their units being "\", "$M^1L^{-1}T^{-2}$", and "$M^1L^{-1}T^{-2}$" respectively, where M, L, and T represent mass, length, and time, respectively. In addition, the inputs also include the unit required at current position, i.e., "$M^1L^{-1}T^{-2}$", and the minimum number of tokens needed to complete the expression, which is "1" at current position. In the actual execution process of the algorithm, all tokens are assigned unique numbers as identifiers, and units are also represented in vector form; for instance, the unit of token "$p_a$" is described as {1, -1, -2}. After receiving its input, each RNN cell outputs a probability distribution for all potential tokens at the corresponding position through the softmax activation function. By applying a multiplier of 0 or 1 to each token, the algorithm can enforce any pre-established constraints, such as unit constraints, nesting constraints, etc.

To demonstrate the implementation of unit constraints, let's consider the prediction process for the fifth token in the expression shown in Fig. 1, assuming that the first four operators {+, $p_a$, ×, $\rho$} in the expression are already accurately predicted. When unit constraints are applied to the expression, the algorithm initially determines the unit required at the current position. It is found that the required unit is "$M^0L^2T^{-2}$", indicating that tokens such as "$p_a$", "$\rho$", "$g$", and "$h$" from the token library are not possible candidates for this position due to unit mismatch. Consequently, a mask vector is generated for all tokens, wherein positions corresponding to ineligible tokens are assigned a value of 0, while the rest are set to 1. This mask vector is then multiplied by the probability distribution vector produced by the RNN cell, effectively reducing the probabilities of the ineligible tokens to 0. This mechanism allows for the integration of any predefined constraints, such as unit constraints, into the prediction process, ensuring that the generated expression adheres to these constraints.

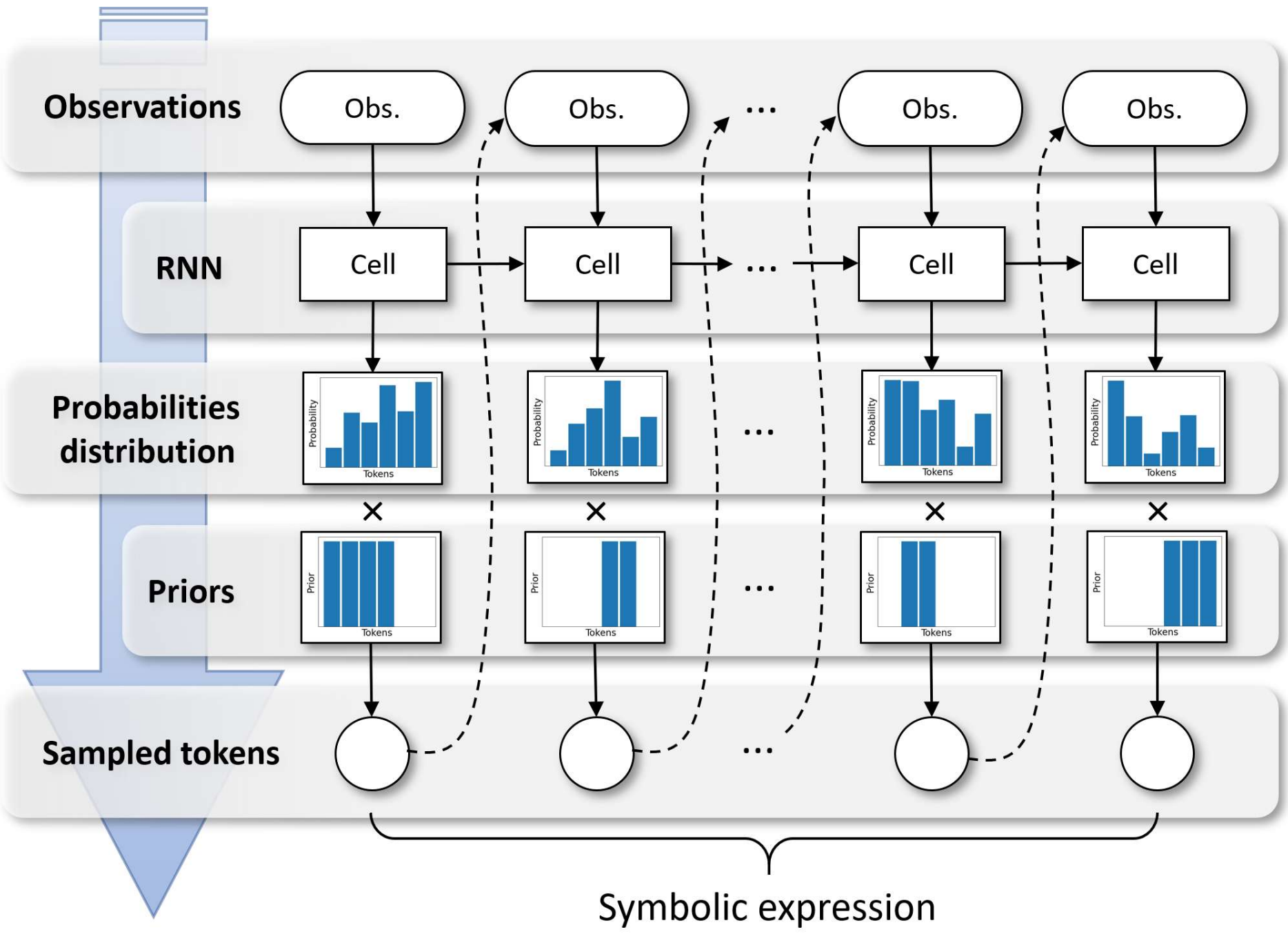

**Fig. 2** Diagram of the Φ-SO Symbolic regression algorithm architecture

Upon the application of prior constraints, the algorithm selects the tokens with the highest probability as the sampling tokens. Eventually, these sampled tokens are positioned to formulate the symbolic expression as output, following the reverse process in Fig. 1 (c→b→a).

To learn knowledge from data, the RNN needs to be trained via the backward propagation process. Typically, Neural networks are trained using the gradient descent (GD) method. However, in the context of symbolic regression problems, the non-differentiability of the cost function with respect to symbol selection prevents the direct application of the GD method for training network. Therefore, the Φ-SO algorithm employs a reinforcement learning strategy to train the network and uses a risk-seeking policy to optimize the performance of symbolic regression. The risk-seeking policy calculates the loss value based on a subset of candidate expressions with the best fitness (i.e., reward values) in a batch, meaning it only learns from the "best-performing" subset of expressions. This approach aligns with the objective of symbolic regression, which seeks to find the best expression that most closely approximates the target, rather than maximizing the average predictive accuracy of all expressions. The effectiveness of the risk-seeking strategy in symbolic regression tasks has been well-documented [34]. For detailed information about the Φ-SO symbolic regression algorithm and the risk-seeking strategy, please refer to the original papers by Tenachi et al. [27] and Petersen et al. [34]

**2. Unit-constrained symbolic regression turbulence modeling**

Based on the Φ-SO symbolic regression algorithm, a unit-constrained turbulence modeling framework using symbolic regression is established, as illustrated in Fig. 3. The proposed framework comprises the following four steps:

**a. Acquisition of baseline flow field and high-precision turbulence data**

Conducting RANS simulation with a baseline LEVM model to acquire the baseline flow field. Subsequently, obtaining high-precision turbulence data through experiments, DNS, or LES simulations, or by accessing existing open-source high-precision turbulence datasets.

**b. Calculation of mean flow features and Reynolds stress deviation**

Using the mesh cell centers from the RANS simulation as sampling points to extract mean flow features, denoted as $q_i$ at each point, where $i$=1, 2, 3, …, $n$, and $n$ is the number of pre-selected mean flow features. The Reynolds stress tensor $\boldsymbol{\tau}_{\text{RANS}}$ at each sampling point is extracted from the baseline flow field. This tensor is then compared with the high-precision Reynolds stress data $\boldsymbol{\tau}_{\text{hp}}$ obtained from a high-accuracy turbulence dataset, to calculate the Reynolds stress deviation tensor $\Delta\boldsymbol{\tau} = \boldsymbol{\tau}_{\text{hp}} - \boldsymbol{\tau}_{\text{RANS}}$ at each sampling point.

**c. Derivation of Reynolds stress deviation expression via symbolic regression**

Using the mean flow features, $q_i$, obtained in the preceding step as input variables, and the Reynolds stress deviation tensor, $\Delta\boldsymbol{\tau}$, as the target variable. Using the unit-constrained symbolic regression algorithm introduced earlier to derive explicit equations linking each component of the Reynolds stress deviation to the mean flow features. As elucidated in Section 2.1, for statistically two-dimensional turbulence, the deviation tensor has four independent, non-zero components, requiring the establishment of four scalar algebraic equations. For three-dimensional turbulence, the deviation tensor has six independent, non-zero components, requiring the establishment of six scalar algebraic equations.

Upon deriving the equations for Reynolds stress deviation $\Delta\boldsymbol{\tau}(q_i)$, these equations are integrated as correction terms into the constitutive equations of the original baseline turbulence model, as depicted in Eq. 7. This integration results in the formulation of a new, learned nonlinear eddy vorticity turbulence model. The methodologies for calculating turbulent viscosity $\mu_t$, turbulent kinetic energy $k$, and other turbulence quantities remain aligned with those of the baseline turbulence model.

$$\boldsymbol{\tau} = \frac{2}{3}k\delta_{ij} - 2\nu_t\mathbf{S} + \boldsymbol{\Delta\tau}(q_i) \tag{7}$$

**d. Numerical simulation using the learned turbulence model**

Finally, the learned turbulence model is implemented within the RANS solver, enabling the generation of an improved turbulence field through a new round of numerical simulation. To improve the convergence of the learned model, the baseline flow field obtained in the first step is used as the initial field for the numerical simulations with the learned model.

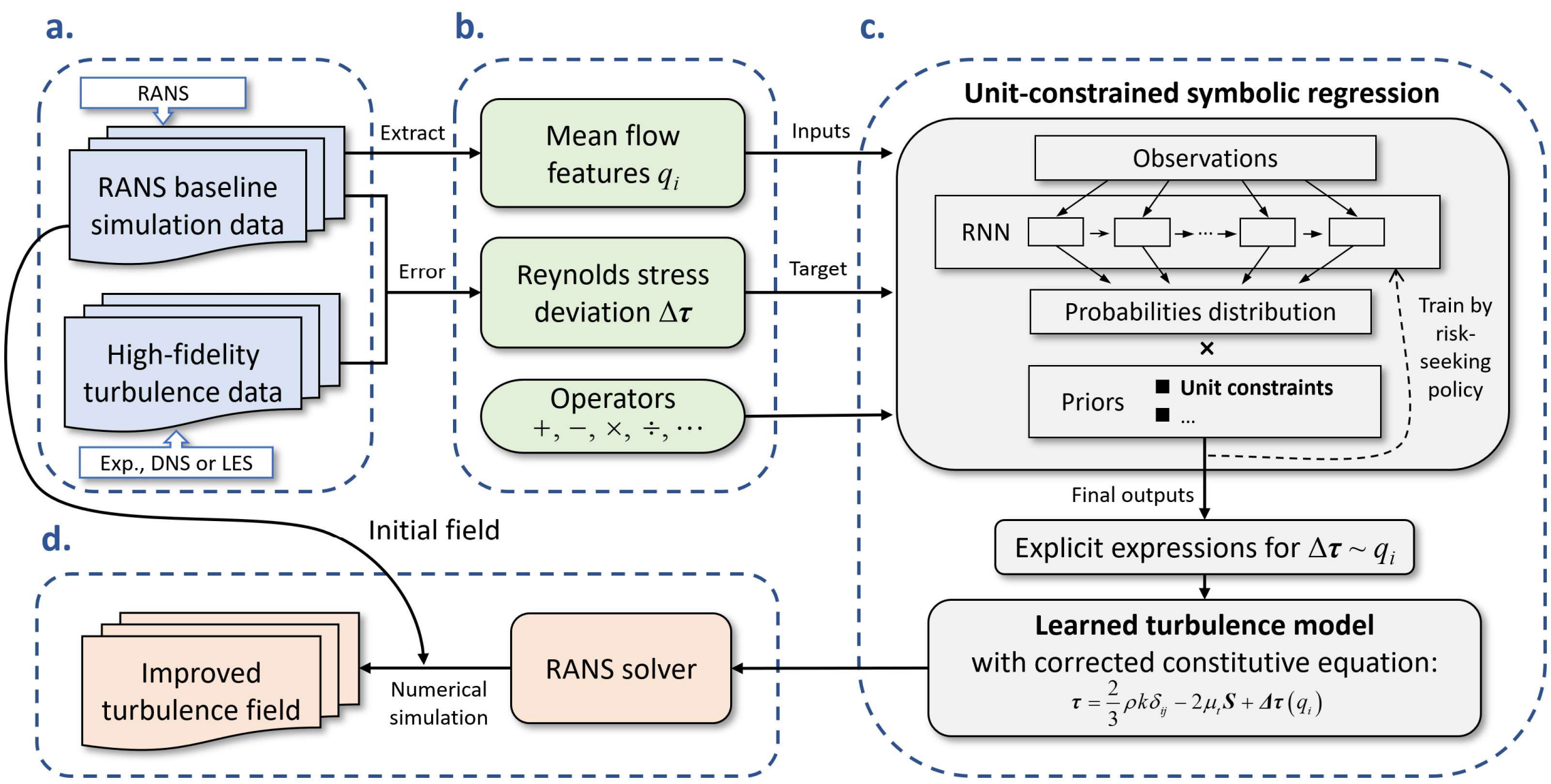

**Fig. 3** A sketch map of the symbolic regression turbulence modeling framework with unit constraint

## 2.3. Numerical method

Numerical simulations in this study are conducted using the open-source software OpenFOAM. The incompressible solver simpleFoam is used to perform steady-state simulations. The convective terms within the governing equation are solved using a second-order upwind scheme, whereas the remaining terms are solved using a second-order central difference scheme. The standard $k$-$\varepsilon$ model proposed by B.E. Launder and D.B. Spalding is selected as the baseline turbulence model. A detailed introduction to the model is provided in Section 2.1 and Reference [29].

# 3. Case Setup

Section 2 introduced the proposed unit-constrained turbulence modeling framework using symbolic regression. To evaluate the efficacy of the proposed framework in large separated turbulence, the subsequent sections of this paper employ the flow over periodic hills as a case study to validate the modeling capabilities of the proposed framework.

This section introduces the basic characteristics of the flow over periodic hills, the selection for input features, the hyperparameters, and the mesh configuration.

### 3.1. Flow over periodic hills

The flow over two-dimensional periodic hills is a commonly used validation case in studying separated turbulence modeling. Xiao et al. [35] presented a series of periodic hills cases with parameterized geometries and provided corresponding DNS datasets. These datasets have been widely used in data-driven turbulence modeling research for separated flows [2, 12, 19, 22].

Figure 4 provides a schematic diagram of the periodic hills with parameterized geometries. The left and right boundaries are subject to periodic boundary conditions, and the top and bottom boundaries are no-slip wall boundaries. The flow enters from the left inlet along the $x$-direction. Upon encountering the first hill on the leeward side, an adverse pressure gradient prompts flow separation, leading to the formation of a recirculation zone. The separated flow reattaches downstream, flows over the windward side of the right second hill, and finally exits through the right boundary. For the periodic hills with parameterized geometries, the total height $L_y$ normalized by the hill height $H$, $L_y/H$, is fixed at 3.036. The total length of the domain along the $x$-direction, $L_x$, is controlled by the steepness parameter $\alpha$. As $\alpha$ increases, $L_x$ will also increase, which in turn stretches the bottom wall along $x$-axis and reduces the slope of the hill. The height of hills $H$ and the total height of the domain $L_y$ do not change with variations in $\alpha$. The equations defining the shape of the hill and the variation of $L_x$ with parameter $\alpha$ are detailed in Appendix A.

In this study, the flow over periodic hills for $\alpha$=0.8 is used as training case for modeling, while $\alpha$=1.0 and 1.2 are used as testing cases. Additionally, the flow over a backward-facing step serves as an extra testing case to validate the learned generalization ability of the learned model in scenarios that differ more significantly from the training conditions, which will be discussed in Section IV.

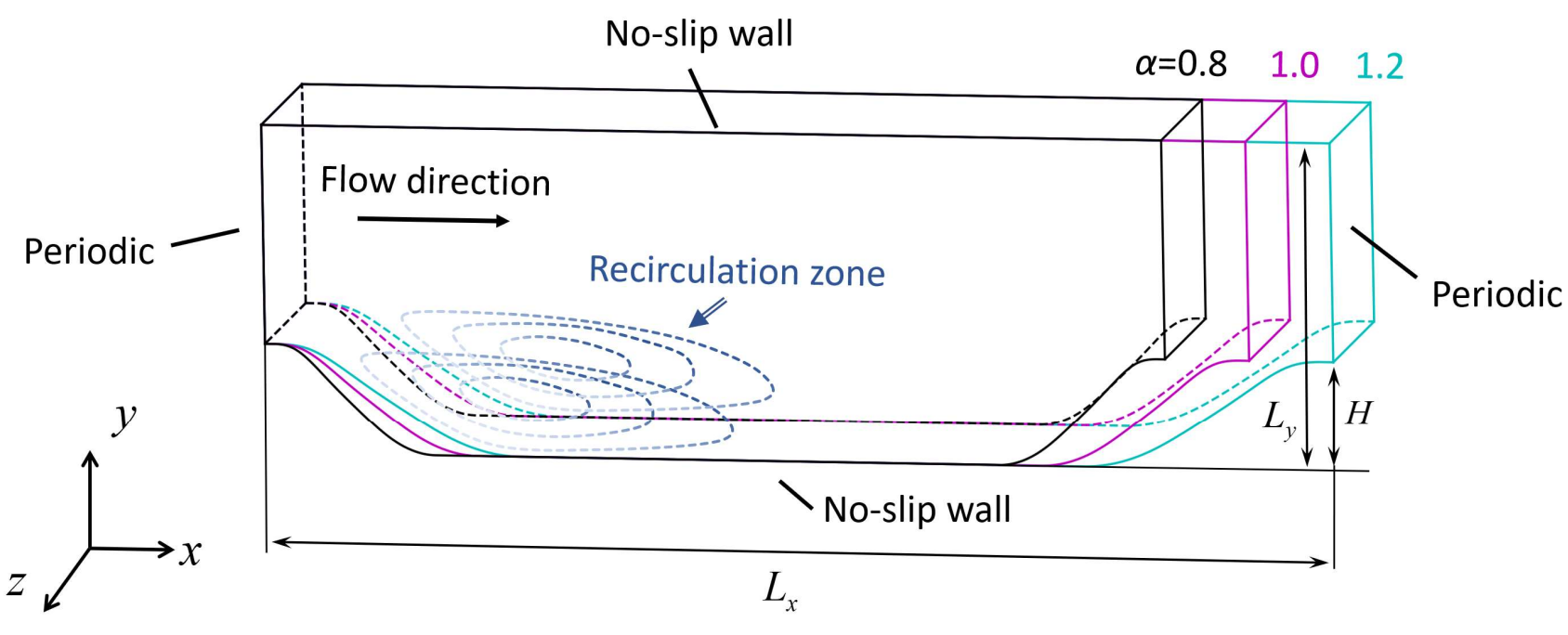

**Fig. 4** A sketch of the flow over periodic hills with parameterized geometries

### 3.2. Mean flow features

The objective of symbolic regression turbulence modeling is to establish explicit equations linking mean flow features $q_i$ to Reynolds stress deviations $\Delta\boldsymbol{\tau}$. Therefore, the selection of mean flow features will directly influence the efficacy of modeling. Considering equation solvability, computational efficiency, and physical significance, the selection of mean flow features should follow these principles:

(1) The features must pertain solely to the mean flow or be derivable from existing turbulence transport equations, ensuring the resultant equations are closed and do not impose additional computational demands.

(2) The features should be able to reflect the characteristics of turbulent flow and have clear physical meanings, ensuring that the results of symbolic regression are interpretable.

(3) The features should satisfy invariance under coordinate transformations, such as rotations.

Based on the principles above, this study selects eight mean flow features proposed by Wang et al. [2] as input features, denoted respectively by $q_1, q_2, \cdots, q_8$. The expressions and physical meanings of these eight features are shown in Table 1. Among them, $q_1$ represents the Q criterion, a parameter commonly used in CFD for identifying flow structures; $q_2$ represents turbulent kinetic energy, indicating the isotropic part of the Reynolds stress tensor, which can be derived from the existing $k$-equation (see Eq. 4); $q_3$ represents the turbulence Reynolds number based on wall distance, which is an important parameter for distinguishing between boundary layers and shear flows; $q_4$ represents the pressure gradient along the streamline; $q_5$ represents the time scale of turbulence, which can be calculated from the existing $k$-equation (see Eq. 4) and $\varepsilon$-equation (see Eq. 5); $q_6$ represents the normal stress part of pressure; $q_7$ represents a measure of the deviation in orthogonality between velocity and its gradient, characterizing the deviation between the flow and parallel shear flows; $q_8$ represents the convection of turbulent kinetic energy.

In addition to the eight mean flow features above, the token library used in this paper also includes five scalar operator tokens and one non-dimensional free constant: $\{+, -, \times, \div, (\cdot)^2, c\}$, where $(\cdot)^2$ denotes the square operator, and $c$ represents the free constant, a total of 14 operators. It is worth mentioning that in existing data-driven turbulence modeling research, Reynolds stresses and input features are often made non-dimensional to avoid non-physical issues arising from unit inconsistencies. In this study, since unit information is an essential constraint for symbolic regression, the features above are not been nondimensionalized.

**Table 1** Mean flow features as symbolic regression input

| Feature | Expression | Physical meanings |
|---|---|---|
| $q_1$ | $\frac{1}{2}\left(\lVert\mathbf{R}\rVert^2-\lVert\mathbf{S}\rVert^2\right)$ | Q criterion |
| $q_2$ | $k$ | Turbulent kinetic energy |
| $q_3$ | $\min\left(\frac{\sqrt{k}d}{50\nu},2\right)$ | Turbulent Reynolds number based on wall distance |
| $q_4$ | $U_k\frac{\partial p}{\partial x_k}$ | Pressure gradient along the streamline |
| $q_5$ | $\frac{k}{\varepsilon}$ | Turbulence time scale |
| $q_6$ | $\sqrt{\frac{\partial p}{\partial x_i}\frac{\partial p}{\partial x_i}}$ | Pressure normal stress |
| $q_7$ | $\left\lvert U_iU_j\frac{\partial U_i}{\partial x_j}\right\rvert$ | Non-orthogonality between velocity and its gradient |
| $q_8$ | $U_i\frac{\partial k}{\partial x_i}$ | Turbulent kinetic energy convection |

### 3.3. Learning parameters

This study employs a risk-seeking strategy to facilitate the learning of the RNN network. The ADAM optimizer is utilized to update network parameters. The optimization of the free constant term in the learned expression is carried out through an embedded inner loop using the LBFGS method. In the risk-seeking gradient policy, the fitness of an expression is measured by the closeness of the calculated result of the learned expression to the target value. The fitness of an expression is quantified using a reward value, where a higher value indicates a smaller error between the symbolic regression result and the true value. The calculation of the reward value is given by:

$$R=\frac{1}{1+\mathrm{NRMSE}} \tag{8}$$

where NRMSE stands for Normalized Root Mean Square Error, and its calculation method is as follows:

$$\mathrm{NRMSE}=\frac{1}{\sigma_{\Delta\tau}}\sqrt{\frac{1}{N}\sum\left(\Delta\tau-f(q_i)\right)^2} \tag{9}$$

where $\sigma_{\Delta\tau}$ is the standard deviation of the target values $\Delta\tau$, $N$ is the size of the dataset, and $f(q_i)$ is the result calculated by learned expression.

The configuration of hyperparameters plays a crucial role in efficiently executing symbolic regression tasks. In this study, each RNN cell consists of 1 hidden layer with 128 neurons; the neural network is trained for 60 epochs, with a batch size of 1000; the risk factor, representing the proportion of preferred expressions selected in the risk-seeking policy among all expressions, is set to 5% in this study; the learning rate is set to 0.0001. To reduce the time required for model training and CFD computations, the maximum length of predicted expressions is set to 15. The initial values of the free constants are set to 1. During free constant optimization, the loss is computed using the NRMSE of the expression. The optimization is performed for 15 steps with a tolerance of $10^{-8}$. The optimization process terminated when the loss fell below the tolerance or the maximum number of steps is reached. A detailed overview of the hyperparameter settings is provided in Table 2. All the hyperparameters underwent meticulous tuning before training to achieve the optimal balance between model prediction accuracy and training cost.

**Table 2** Hyper-parameters

| Hyper-parameter | Value |
|---|---|
| RNN architecture | 128×1 |
| Num. of epoch | 200 |
| Risk factor | 5% |
| Batch size | 10000 |
| Entropy weight | 0.005 |
| Gamma decay | 0.7 |
| Max length | 15 |
| Learning rate | 0.0001 |
| Initial value of free constants | 1 |
| Num. of LBFGS optimization steps | 15 |
| Tolerance of free constant optimization | $10^{-8}$ |

## 3.4. Mesh and mesh independence study

In the numerical simulation part of this paper, a structured mesh, as shown in Fig. 5, is used to discretize the computational domain. The mesh is appropriately refined near the walls and corners to capture the boundary and recirculation region accurately. To verify the mesh independence of the numerical results, three sets of meshes, namely Mesh A, Mesh B, and Mesh C, from sparse to dense, are generated for mesh independence validation. Table 3 provides the distribution of mesh points along the streamwise ($x$-axis), wall-normal ($y$-axis), and spanwise ($z$-axis) as well as the non-dimensional first-layer mesh height $\Delta y/H$ for the three mesh sets. Since this study focuses on two-dimensional turbulent flow, the number of mesh points in the spanwise ($z$-axis) for all three mesh sets are set to 2.

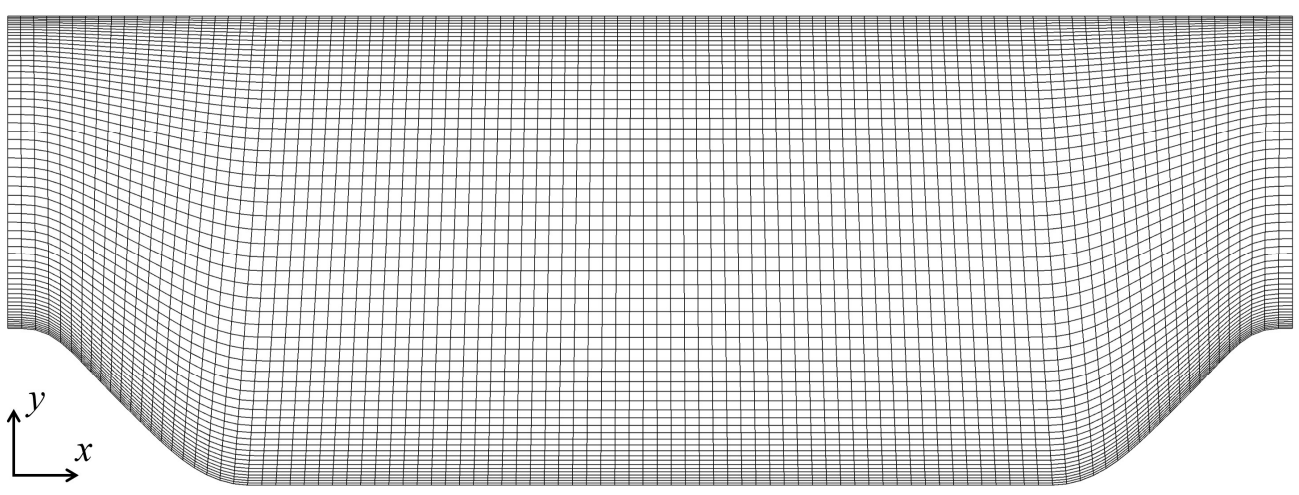

**Fig. 5** Mesh of computation domain

**Table 3** Summary of mesh resolution along different directions

| Mesh | Streamwise (x-direction) | Wall-normal (y-direction) | Spanwise (z-direction) | $\Delta y/H$ |
|---|---|---|---|---|
| Mesh A | 100 | 60 | 2 | $9\times10^{-3}$ |
| Mesh B | 140 | 80 | 2 | $6\times10^{-3}$ |
| Mesh C | 160 | 100 | 2 | $3\times10^{-3}$ |

Figure 6 presents the streamwise velocity (normalized by the bulk velocity $U_b$ at the hill crest, where $U_b = 0.028$m/s) profiles for flow over periodic hills at $x/H$=1.5 for each mesh, as obtained by the standard *k*-*ε* model under the conditions of $\alpha$=0.8 and Re=5600. It can be observed that the streamwise velocity distributions obtained from these three meshes are approximately the same. Upon closer examination of the curves, a relatively significant discrepancy is observed between the velocity distributions obtained from Mesh A and Mesh B, while the results from Mesh B and Mesh C exhibit good agreement. Therefore, Mesh B and Mesh C meet the requirements of mesh independence. Considering both computational efficiency and accuracy, Mesh B is selected to compute the baseline flow field and subsequent validation of the learned models.

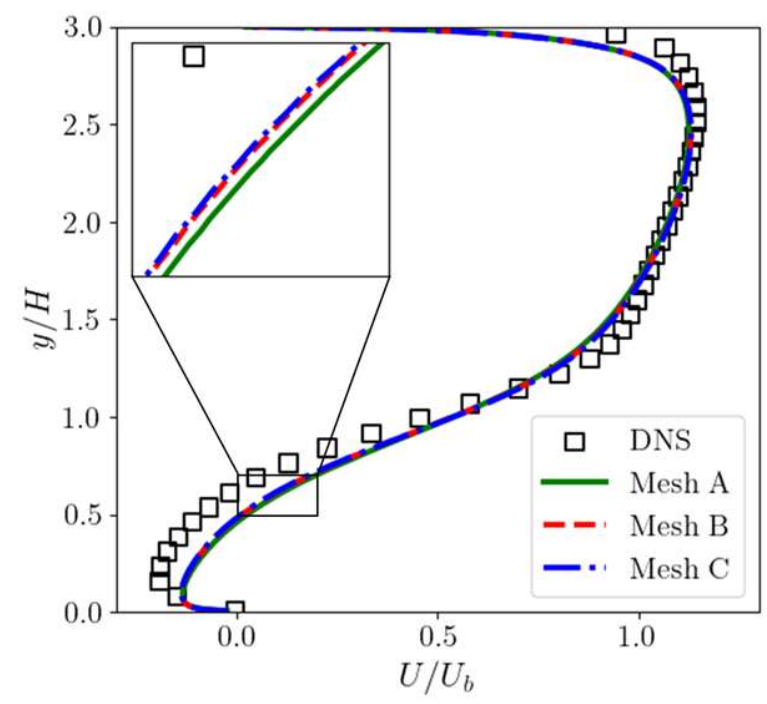

**Fig. 6** Normalized streamwise velocity profile at $x/H$=1.5 calculated by three different meshes ($\alpha$=0.8, Re=5600)

## 4. Results

To verify the effectiveness of the proposed framework in large separated flow, this section first uses the standard *k*-*ε* model to perform numerical simulation of the flow over periodic hills and obtain the baseline flow field. Then, the eight mean flow features given in Section 3 and the Reynolds stresses are derived. By comparing to the Reynolds stresses extracted from the high-fidelity DNS turbulence dataset, the Reynolds stress deviations are calculated. The Reynolds stress deviations and the mean flow features are used as inputs for the symbolic regression learning to obtain explicit equations for the Reynolds stress deviation. These equations are then embedded as a correction term into the constitutive equation of the original turbulence model to establish a learned turbulence model. Finally, the performance of the learned model is validated through numerical simulations under both the training conditions and scenarios that differ from the training data.

### 4.1. Learning result

In this study, the flow over periodic hills under the condition of $\alpha$=0.8 was used as the training case. The flow Reynolds number was 5600, which is based on the wall height $H$ and the bulk velocity $U_b$ at the hill crest. The numerical methods and mesh described in Section 3 were employed, and the standard $k$-$\varepsilon$ model was used to compute the baseline flow field. The computation was performed for 20,000 iterations, ensuring convergence by examining the residuals.

The Reynolds stresses obtained from the baseline flow field, along with the flow characteristic quantities and DNS Reynolds stress data, were used as inputs for the symbolic regression algorithm. After training for 200 epochs, explicit equations for the four independent, non-zero components of the Reynolds stress tensor discrepancies were obtained. The hyperparameter configuration used for the training is detailed in Table 2. The training was conducted on a computer with an AMD 7950X processor and 64GB of RAM. Each of the four Reynolds stress discrepancy components was trained for 200 epochs, with the total training time amounting to approximately 7.5 minutes per CPU core.

The variation of the reward value of the best expression in a batch with the number of epochs during the learning is shown in Fig. 7. It is obvious that the reward values corresponding to the four Reynolds stress discrepancy components initially increase rapidly with the number of epochs and then stabilize, indicating that the symbolic regression training has converged within the selected number of epochs. The final expressions were selected from the Pareto front characterized by expression complexity and reward through a comprehensive evaluation of their convergence behavior and error characteristics.

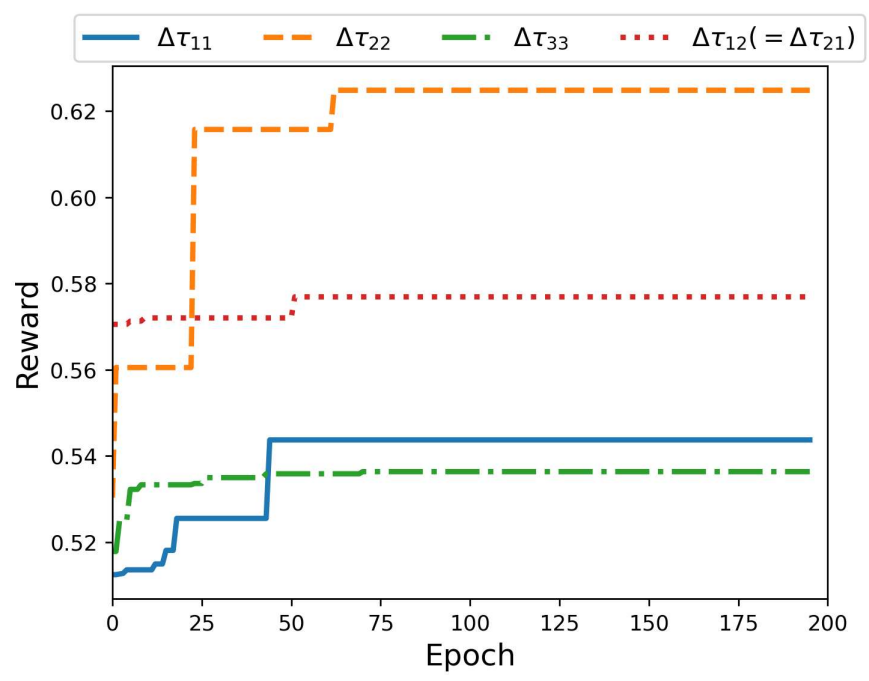

**Fig. 7** The variation of reward values with the epochs during the learning process

The explicit equations for the four components of the Reynolds stress discrepancy tensor obtained through symbolic regression are as follows:

$$\begin{cases} \Delta\tau_{11} = c_1 q_2 q_3^2, \; c_1 = 0.0454 \\ \Delta\tau_{22} = -\dfrac{q_2}{q_3^2 - q_3 + c_2}, \; c_2 = 2.2139 \\ \Delta\tau_{33} = q_2 q_3^2\left(c_3 q_3 + 1\right), \; c_3 = \text{-}0.5072 \\ \Delta\tau_{12}(=\Delta\tau_{21}) = \dfrac{q_5 q_8}{q_3^2 + c_4}, \; c_4 = 2.5347 \end{cases} \tag{10}$$

where $c_1$ to $c_4$ represent the free constants in the Reynolds stress discrepancy expressions.

Eq. 10 shows that the trained Reynolds stress discrepancies are solely related to four mean flow features: turbulent kinetic energy $q_2$, turbulent Reynolds number $q_3$, turbulent time scale $q_5$, and turbulent kinetic energy convection $q_8$. Specifically, the three normal stress components of the Reynolds stress discrepancy, $\Delta\tau_{11}$, $\Delta\tau_{22}$, and $\Delta\tau_{33}$, are only functions of $q_2$ and $q_3$, while the shear stress component $\Delta\tau_{12}$ is determined by $q_3$, $q_5$, and $q_8$. By examining the units on both sides of the equations and the addition or subtraction, it is confirmed that the obtained expressions adhere to the principle of unit consistency.

Figure 8 presents the spatial distributions of the four mean flow features (first row) and the four non-dimensional Reynolds stress deviation components derived from Eq. 10 (second row) for the periodic hill case at $\alpha = 0.8$. For $\Delta\tau_{11}$, by definition both $q_2$ and $q_3$ are positive, ensuring that $\Delta\tau_{11}$ is positive. Specifically, $q_3$ attains a constant value of 2 away from the wall but decreases with decreasing wall-normal distance, reaching zero at the wall. Consequently, in the outer region $\Delta\tau_{11}$ is dominated by the turbulent kinetic energy and follows a nearly identical spatial distribution (see column 1 in Fig. 8). In the near-wall region, the quadratic dependence on $q_3$ accelerates the convergence of $\Delta\tau_{11}$ to zero. $\Delta\tau_{22}$ is negative and, similar to $\Delta\tau_{11}$, is dominated by the turbulent kinetic energy in the outer region. However, the presence of the denominator causes a slight decrease near the wall. The structure of $\Delta\tau_{33}$ is particularly interesting. The pre-factor $q_2 q_3^2$ shares the same structure as $\Delta\tau_{11}$, but the term within the parentheses acts as a switch function. In regions far from the wall where $q_3$= 2, $\Delta\tau_{33}$ is nearly zero; as the wall is approached, the inner term becomes positive, thereby confining the correction to the near-wall region. Compared with the three normal stress components, $\Delta\tau_{12}$ exhibits a more complex dependence, being simultaneously affected by $q_5$, $q_8$, and $q_3$. In the outer region, its behavior is controlled by $q_5$ and $q_8$—with $q_8$ playing the dominant role, as evident in column 4 of Fig. 8.

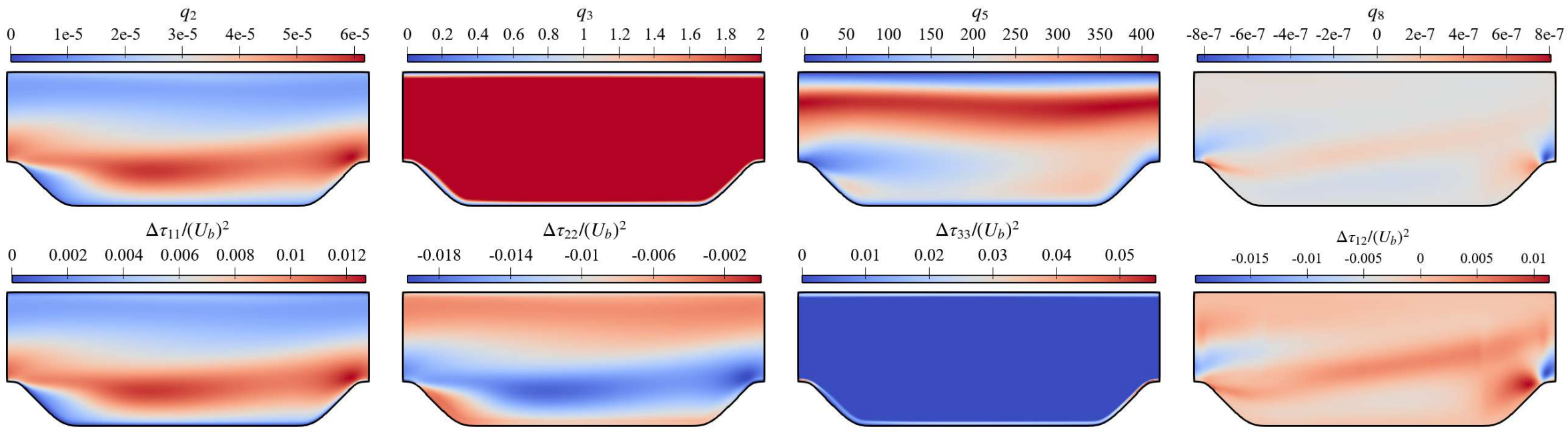

**Fig. 8** Distribution of mean flow features and non-dimensional Reynolds stress deviations (periodic hills, $\alpha$=0.8)

A sensitivity analysis of the symbolic regression results with respect to the RANS mesh resolution was conducted. Training datasets were constructed using RANS data obtained from three meshes of varying resolutions as listed in Table 3, and symbolic regression was performed.

The results show that all datasets yielded the same structural form of the Reynolds stress discrepancy expression, with only slight variations in the values of the free constants. Table 4 presents the free constants in each expression obtained from different meshes. The mean relative error of the free constants is 4.12% between Mesh A and Mesh B, and 3.89% between Mesh B and Mesh C. These findings indicate that the symbolic regression results are robust to mesh resolution and training data, suggesting that the learned expression in Eq. 10 is not a result of overfitting to a specific dataset.

**Table 4** Free constants in the learned expressions for different meshes

| Mesh | $c_1$ | $c_2$ | $c_3$ | $c_4$ |
|---|---|---|---|---|
| Mesh A | 0.0493 | 2.3503 | -0.5061 | 2.5733 |
| Mesh B | 0.0454 | 2.2139 | -0.5072 | 2.5347 |
| Mesh C | 0.0435 | 2.1209 | -0.5074 | 2.3539 |

By reconstructing the Reynolds stress deviation components given by Eq. 10 into the Reynolds stress tensor, and incorporating it as a correction term into the constitutive equation of the original turbulence model, the modified constitutive equation shown in Eq. 7 is obtained, thereby establishing a new learned turbulence model.

### 4.2. Performance in the training geometry

To validate the predictive accuracy of the learned model for separated turbulence flow, numerical simulations were performed using the learned model on the periodic hills flow ($\alpha$=0.8) with the same geometry as the training case. The baseline flow field was used as the initial field for the simulation. The variation of residuals for the flow variables during simulation are shown in Fig. 9. It can be observed that, prior to 10000 iterations, the residuals for all variables decrease rapidly to below $10^{-9}$. This demonstrates that the learned model exhibits satisfactory convergence properties.

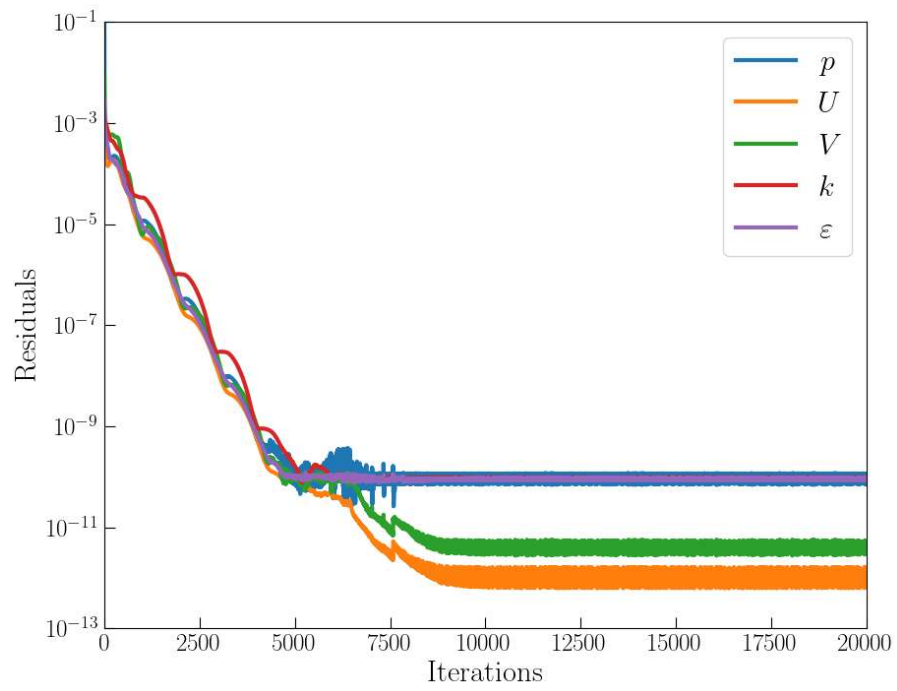

**Fig. 9** Variation of residuals during computation

Figure 10 presents the distributions of the non-dimensional anisotropic Reynolds stress tensor components $\boldsymbol{a}/(U_b)^2$ and their errors for the flow over periodic hills ($\alpha$=0.8, training case) obtained by the standard *k*-*ε* model and learned turbulence model, along with comparisons to DNS results. The first three rows correspond to the results from the *k*-*ε* model, the learned model, and DNS, respectively; the fourth and fifth rows display the absolute errors between the predictions of the *k*-ε model and the learned model compared to the DNS results. Comparing the first row (standard *k*-*ε* model results) and the third row (DNS results), it is evident that the spatial distributions of the anisotropic Reynolds stress components from the standard *k*-*ε* model significantly differ from the DNS results. For component $a_{11}$, the standard *k*-*ε* model underestimates its values in the non-parallel free shear flow region ($y/H$=0.8~1.5) and near the crest of the windward side of the hill ($x/H\approx 8.0$). Notably, the non-parallel free shear flow region is precisely where RANS simulations typically encounter prediction inaccuracies in such flows [36]. For $a_{22}$, the standard *k*-*ε* model overestimates its values in the non-parallel free shear flow region and near the crest of the windward side of the hill. For $a_{33}$, the standard *k*-*ε* model's results are nearly zero across the entire domain, failing to capture its spatial distribution. For the shear stress component $a_{12}$, the predictions from the standard *k*-*ε* model are generally close to the DNS results, but the values are significantly underestimated near the windward side of the hill.

Comparing the second row (learned model results) and the third row (DNS results), it is evident that the learned model significantly improves the prediction accuracy of the anisotropic Reynolds stress components in most regions compared to the standard *k*-*ε* model. For the normal stress component $a_{11}$, the learned model correctly captures the sign and distribution in the non-parallel free shear flow region, correcting the underestimation of $a_{11}$ observed with the *k*-*ε* model on the windward side of the hill. For component $a_{22}$, near the crest of the windward side of the hill, the learned model reduces the overestimation compared to the standard *k*-*ε* model, bringing the values closer to the DNS results. Additionally, the learned model correctly captures the sign and distribution of $a_{22}$ in the non-parallel free shear flow region, although it exhibits some overcorrection. For component $a_{33}$, the learned model captures its distribution near the bottom wall but fails to predict the distribution further away from the wall, and overestimates the values near the upper wall. For component $a_{12}$, the learned model shows improvements over the standard *k*-*ε* model, particularly near the windward side of the hill.

The error distributions presented in the fourth and fifth rows indicate that, for components $a_{11}$ and $a_{12}$, the learned model significantly reduces prediction errors in most regions compared to the *k*-*ε* model. For $a_{22}$, however, the overcorrection by the learned model results in increased errors near the leeward side of the hill and in the non-parallel free shear flow region; nonetheless, the error on the windward side of the hill is significantly reduced. Regarding $a_{12}$, the learned model accurately predicts the distribution in the thin layer near the wall, although the overall improvement in error is not substantial.

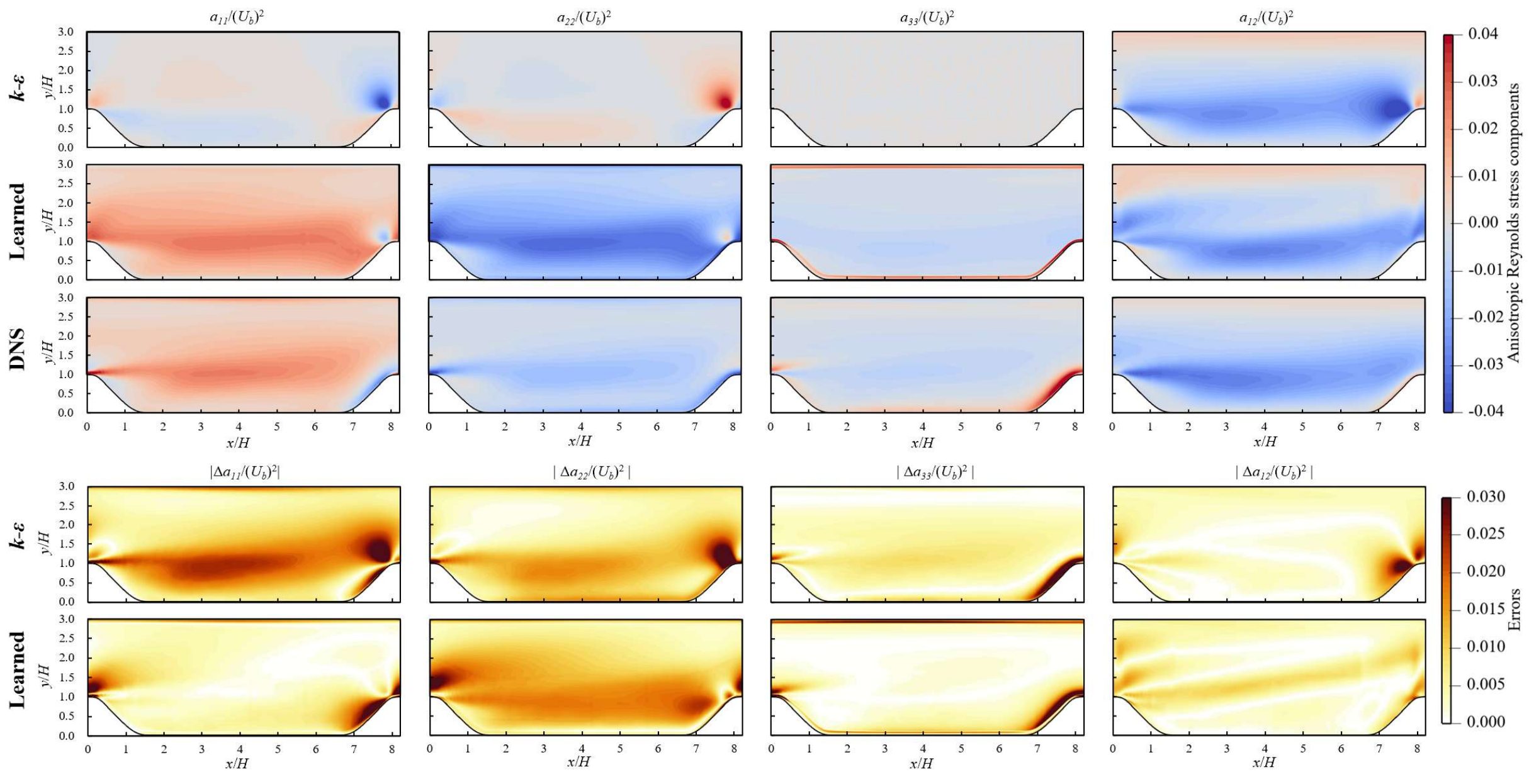

**Fig. 10** Non-dimensional anisotropic Reynolds stress components and errors obtained by the standard *k*-*ε* model and learned model, compared with DNS data (*α*=0.8, training case)

Quantities such as flow velocity are often of greater interest in practical engineering applications. Wu et al. [8] pointed out that in data-driven turbulence modeling studies, even if the Reynolds stresses have high accuracy (errors less than 0.5%), the resulting velocity field can still exhibit significant errors (up to 35%). Therefore, further validation of the learned model's performance in predicting the velocity field is necessary.

Figure 11 presents the non-dimensional streamwise velocity distributions and streamlines for the flow over periodic hills ($\alpha$=0.8, training case) obtained from the standard $k$-$\varepsilon$ model and the learned turbulence model, compared with DNS results. It can be seen that the standard $k$-$\varepsilon$ model, the learned turbulence model, and the DNS all capture the flow separation on the leeward side of the hill ($x/H$=0~4), but there are differences in the shape and size of the recirculation zones obtained by the three methods. Compared to the DNS results, the standard $k$-$\varepsilon$ model predicts a further downstream separation point and a reattachment point that is further upstream, resulting in a significantly smaller recirculation zone than that observed in the DNS results. In contrast, the learned model yields separation and reattachment points, as well as the size of the recirculation zone, that are much closer to the DNS results.

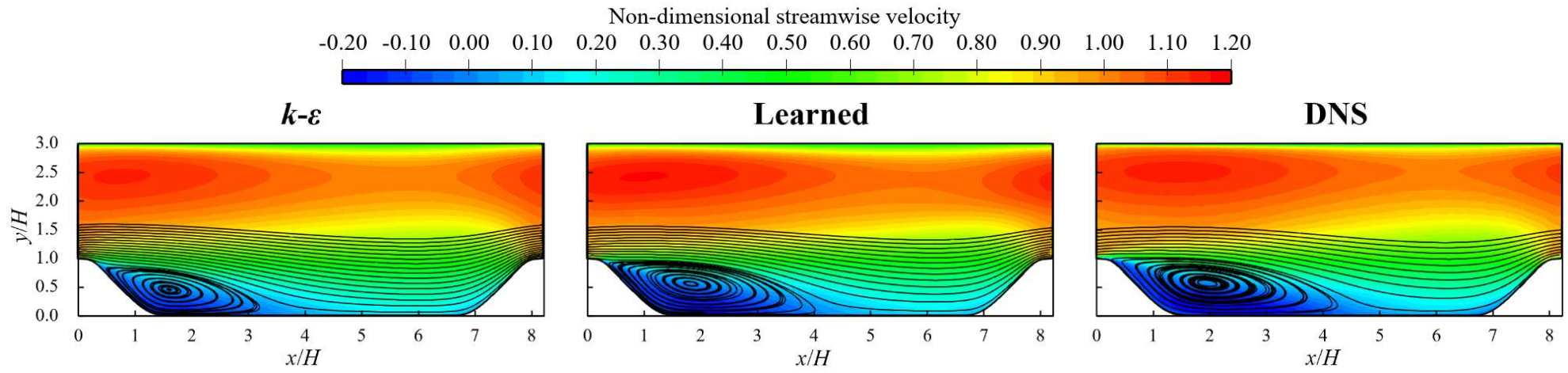

**Fig. 11** Non-dimensional streamwise velocity distribution and streamlines obtained by the standard $k$-$\varepsilon$ model and the learned model, compared with DNS data ($\alpha$=0.8, training case)

The skin friction $C_f$ is a key indicator in the study of separated turbulence modeling, calculated by $C_f = \tau_w / \left(\frac{1}{2}\rho U_b^2\right)$, where $\tau_w$ represents the wall shear stress. Figure 12 presents the distribution of the skin friction in the downstream region of the periodic hills ($\alpha$ = 0.8) from $x/H$=2.5 to $x/H$=5.0. It can be observed that the distribution of the skin friction obtained from the learned model is closer to the DNS results compared to the $k$-$\varepsilon$ model. The flow reattachment point corresponds to the position where the skin friction is zero, and the results indicate that the flow reattachment positions predicted by DNS, the learned model, and the $k$-$\varepsilon$ model are approximately at $x/H\approx$ 5.0, 4.3, and 3.7, respectively. The learned model's prediction of the flow reattachment position is more consistent with the DNS results than that of the $k$-$\varepsilon$ model, which is consistent with the findings shown in Fig 11. Furthermore, in the recirculation region, all three methods predict local peaks in the skin friction; however, both the magnitude of the peak and its corresponding location are closer to the DNS results when using the learned model compared to the $k$-$\varepsilon$ model.

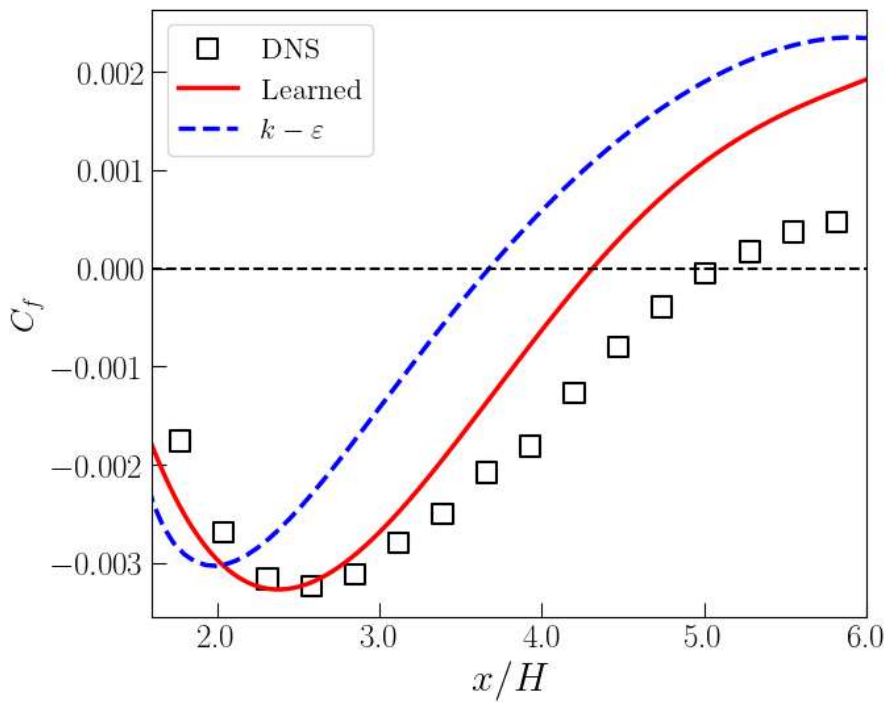

**Fig. 12** Skin friction distribution along the bottom wall obtained by the standard *k-ε* model and the learned model, compared with DNS data (*α*=0.8, training case)

Figure 13 shows the non-dimensional streamwise velocity profiles at different streamwise positions (*x/H*=0.5 to 7.5, with a step size of 1.0) for the flow over periodic hills (*α*=0.8, training case) obtained from the standard *k-ε* model and the learned model, compared with DNS results. As shown in Fig. 13, both the standard *k-ε* model and the learned turbulence model capture the general trend of the streamwise velocity profiles at different streamwise positions. However, in the recirculation zone (*x/H*=0~5.0), the streamwise velocity distributions obtained from the standard *k-ε* model show significant discrepancies compared to the DNS results. Specifically, the standard *k-ε* model underestimates the velocity in the recirculation zone and fails to capture the near-wall recirculation at *x/H*=3.5. In contrast, the learned model accurately captures the streamwise velocity distribution within the recirculation zone. Additionally, as shown in Fig. 10, although the learned model overestimates the anisotropic Reynolds stress component $a_{33}$ near the upper wall, this overestimation does not manifest in the corresponding region of the velocity field.

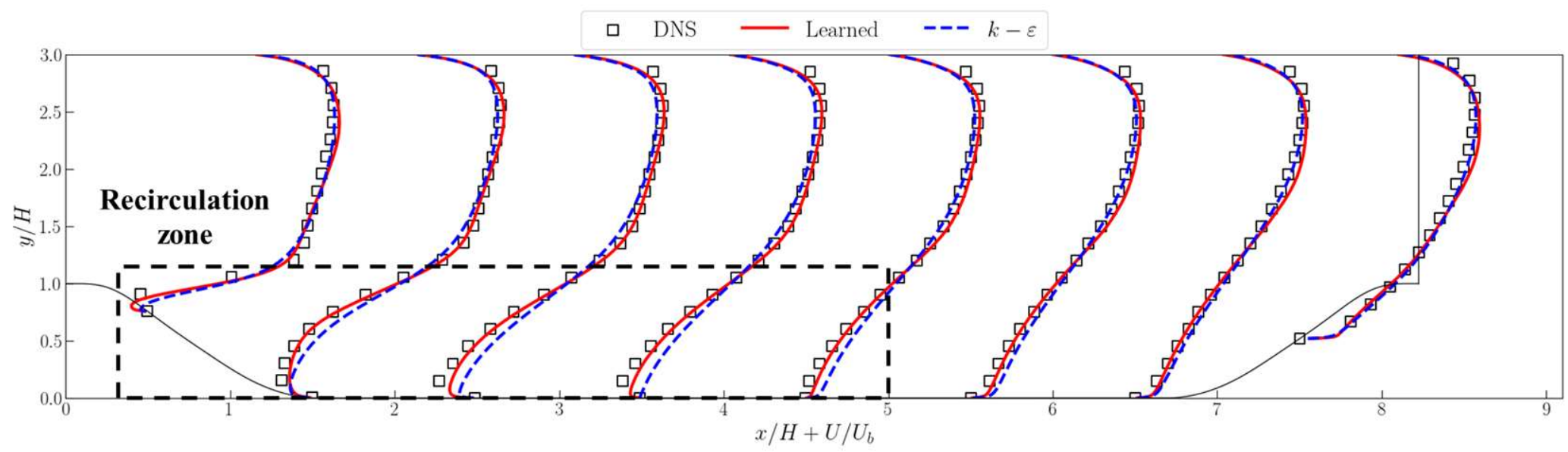

**Fig. 13** Non-dimensional streamwise velocity profile obtained by the standard *k-ε* model and the learned model, compared with DNS data (*α*=0.8, training case)

From the results above, in the training case (*α*=0.8), the learned model derived from unit-constrained symbolic regression turbulence modeling not only achieves higher accuracy in predicting the anisotropic Reynolds stress tensor compared to the standard $k$-$\varepsilon$ model but also demonstrates improved accuracy in predicting the skin friction and the streamwise velocity profiles.

### 4.3. Beyond the training scope

In data-driven turbulence modeling research, a common issue is the insufficient generalization performance of the established models. This means that when the freestream conditions or the geometry of the flow field change, the model's performance can deteriorate significantly compared to its performance under the training conditions, thereby limiting its applicability. To assess the generalization performance of the learned model, three testing cases featuring different geometries or Reynolds numbers were studied: the flow over periodic hills for $\alpha$=1.0 and $\alpha$=1.2, and flow over a backward-facing step. The flow over periodic hills with $\alpha$=1.0 and $\alpha$=1.2 exhibit less steep wall geometries than the training case ($\alpha$=0.8), while maintaining a consistent Reynolds number of 5600. The backward-facing step case, on the other hand, presents a distinctly different geometry and a higher Reynolds number of 36000, which poses a greater challenge for the learned model's generalization ability under varying geometries and flow conditions. The numerical methods used in these testing cases remained consistent with those described in Section 3. The results for the two periodic hills cases are presented first.

Figure 14 shows the distributions of the non-dimensional anisotropic Reynolds stress tensor components and their absolute errors for the flow over the periodic hills ($\alpha$=1.0, non-training case) obtained from the standard $k$-$\varepsilon$ model and the learned model respectively. It can be observed that similar to the training case ($\alpha$=0.8), the prediction errors of the anisotropic Reynolds stress tensor from the standard $k$-$\varepsilon$ model are primarily located in the non-parallel free shear flow region ($y/H$=0.8~1.5) and near the windward side of the hill ($x/H\approx$8.5). However, due to the increased $\alpha$, which results in a more gradual slope of the hill, the adverse pressure gradient is reduced, and the flow separation is weakened. Consequently, the prediction errors of the anisotropic Reynolds stresses from the standard $k$-$\varepsilon$ model are relatively smaller than the training case ($\alpha$=0.8). Compared to the standard $k$-$\varepsilon$ model, the learned model still shows significant improvements in the accuracy of the anisotropic Reynolds stress tensor prediction, with the most notable enhancements in the predictions of the components $a_{11}$, $a_{22}$, and $a_{12}$.

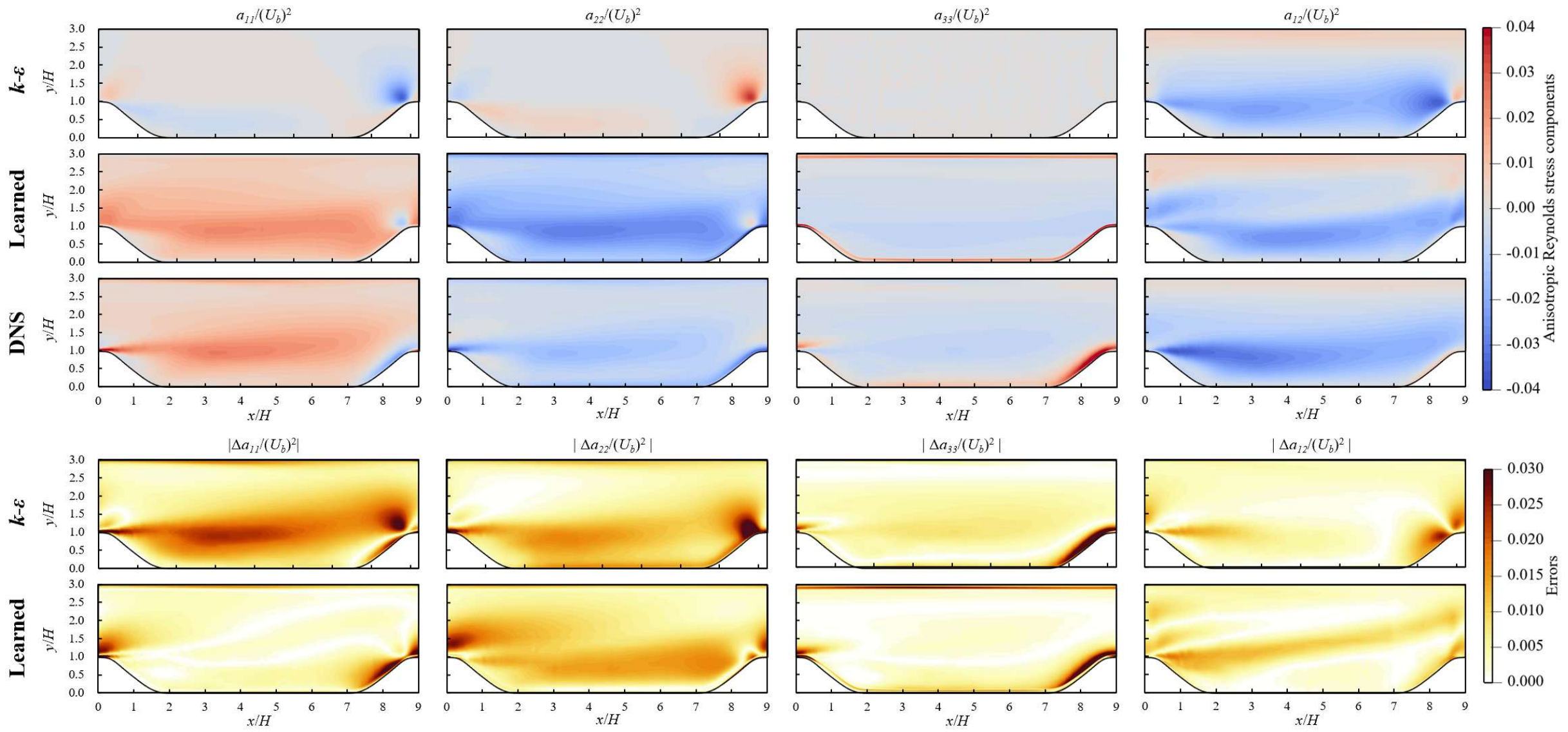

**Fig. 14** Non-dimensional anisotropic Reynolds stress components and errors obtained by the standard *k-ε* model and learned model, compared with DNS data (*α*=1.0, non-training case)

Figure 15 shows the non-dimensional streamwise velocity distributions and streamlines for the flow over the periodic hills (*α*=1.0, non-training case) obtained from the standard *k-ε* model and the learned turbulence model, compared with DNS results. Figure 16 presents the skin friction distribution along the bottom wall obtained from these three methods. As shown in Fig. 15, increasing *α* leads to a more gradual slope of the hill, resulting in a slight reduction in the size of the recirculation zone compared to the *α*=0.8 case. Similar to the case with *α*=0.8. The *k-ε* model significantly underestimates the size of the recirculation zone, and the learned model provides a more accurate prediction of the recirculation zone size. From Fig. 16, it is evident that the skin friction predicted by the learned model closely aligns with the DNS results in both trend and magnitude, significantly outperforming the results of the *k-ε* model. Furthermore, the reattachment location obtained from DNS is approximately located at $x/H\approx4.6$, while the learned model predicts the reattachment location to be around $x/H\approx4.4$. In contrast, the *k-ε* model predicts the reattachment location to be approximately $x/H\approx3.6$. These results indicate that the learned model offers a substantially more accurate prediction of the reattachment location compared to the *k-ε* model.

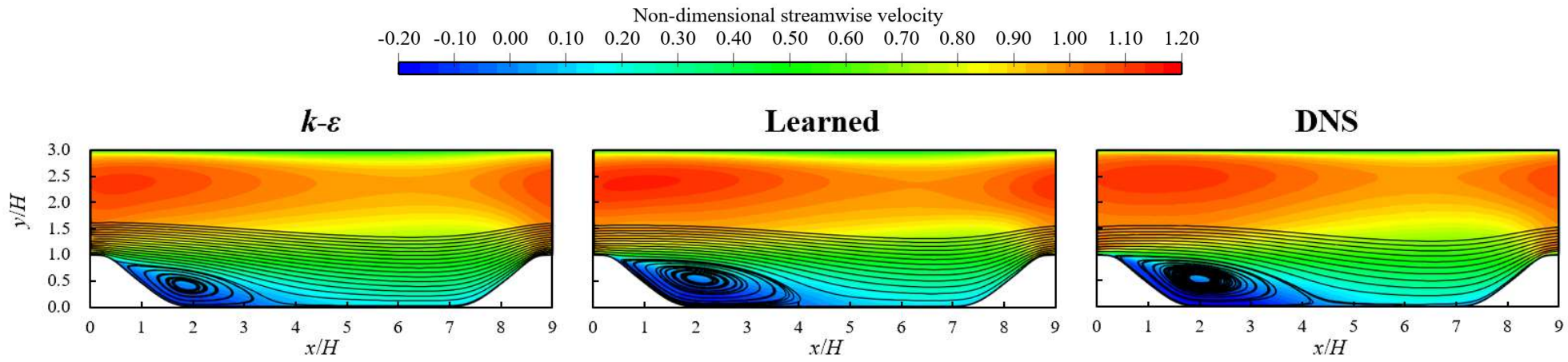

**Fig. 15** Non-dimensional streamwise velocity distribution and streamlines obtained by the standard *k*-*ε* model and the learned model, compared with DNS data (*α*=1.0, non-training case)

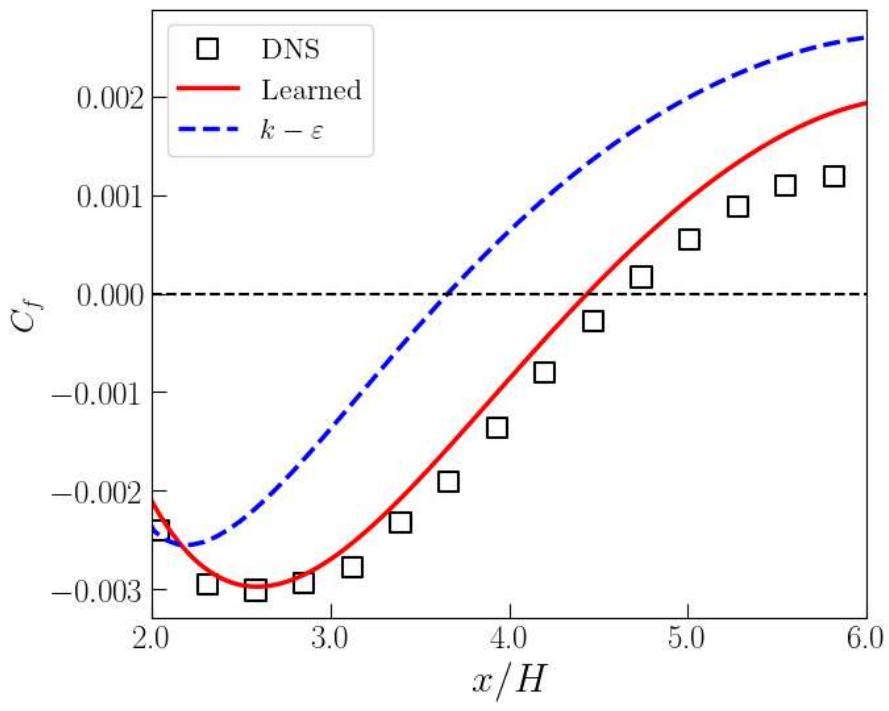

**Fig. 16** Skin friction distribution along the bottom wall obtained by the standard *k*-*ε* model and the learned model, compared with DNS data (*α*=1.0, training case)

Figure 17 presents the non-dimensional streamwise velocity profiles at different streamwise positions (*x/H*=0.5 to 7.5, with a step size of 1.0) for the flow over periodic hills (*α*=1.0, non-training case) obtained from the standard *k*-*ε* model and the learned turbulence model, compared with DNS results. The results demonstrate that the streamwise velocity profiles predicted by the learned model are closer to the DNS results than those from the standard *k*-*ε* model, demonstrating even greater accuracy than observed in the training cases, particularly in the recirculation zone.

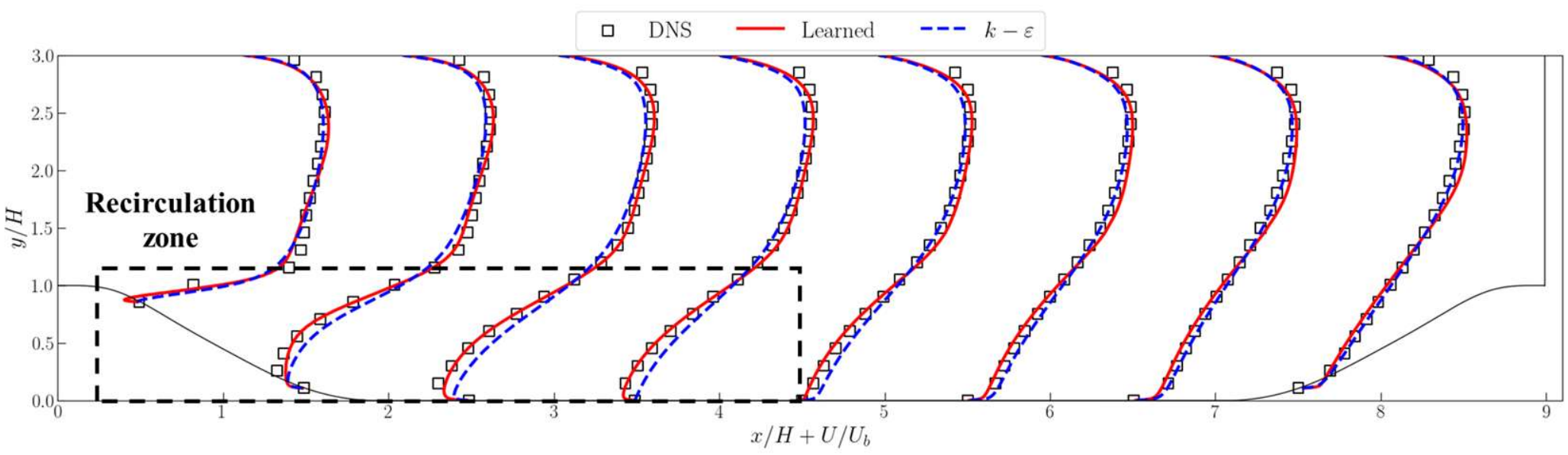

**Fig. 17** Non-dimensional streamwise velocity profile obtained by the standard *k*-*ε* model and the learned model, compared with DNS data (*α*=1.0, non-training case)

Next, the test results for periodic hill flow for *α*=1.2 are discussed. Figure 18 present the distributions of the non-dimensional anisotropic Reynolds stress tensor components and their absolute errors for the flow over periodic hills (*α*=1.2, non-training case) obtained from the standard *k*-*ε* model and the learned turbulence model, respectively.

Notably, even with a further reduction in the steepness of the hill, the learned model demonstrates a smaller error in the distribution of anisotropic Reynolds stress compared to the *k-ε* model.

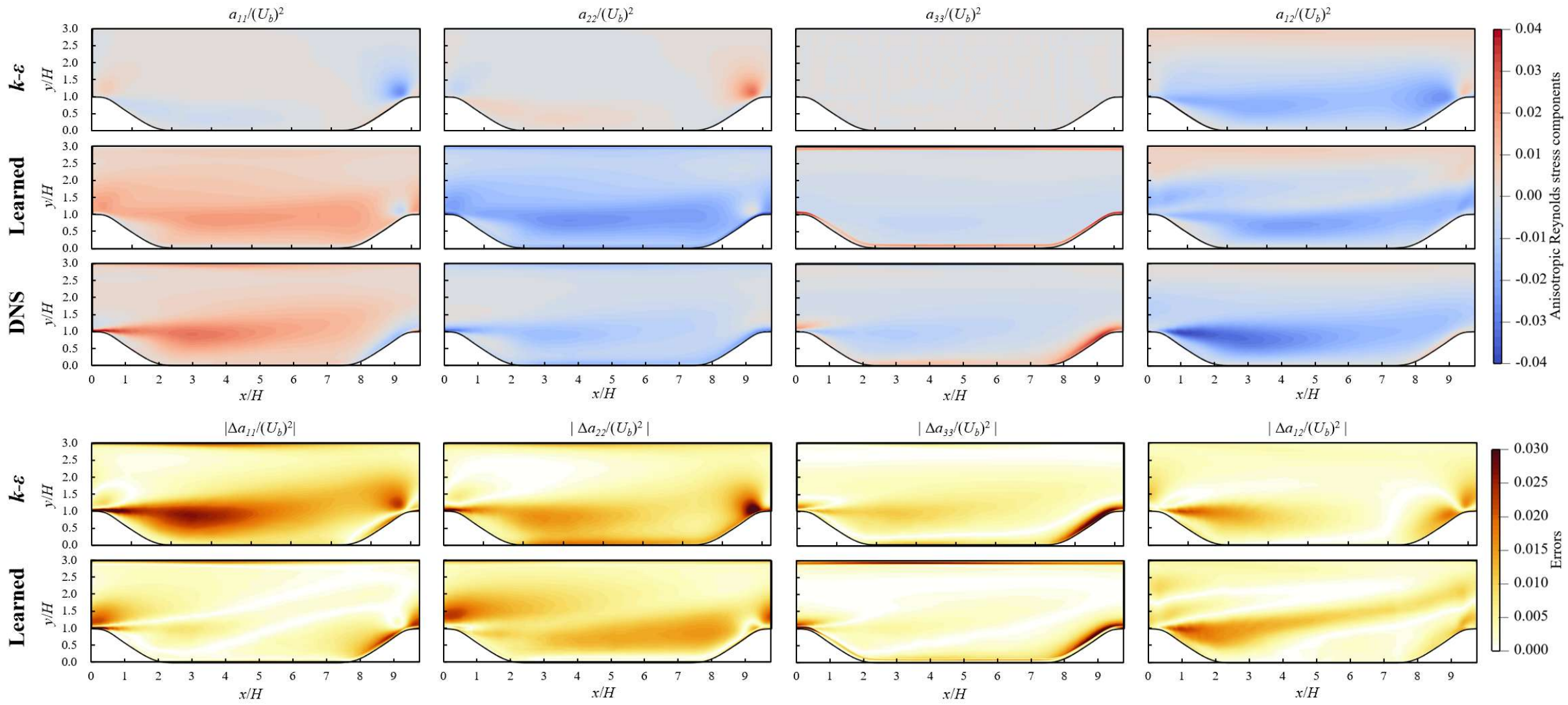

**Fig. 18** Non-dimensional anisotropic Reynolds stress components and errors obtained by the standard *k-ε* model and learned model, compared with DNS data (*α*=1.2, testing case)

Figure 19 shows the non-dimensional streamwise velocity distributions and streamlines for the periodic hills flow (*α*=1.2, non-training case) obtained from the standard *k-ε* model and the learned model, compared with DNS results. Figure 20 presents the skin friction distribution along the bottom wall obtained from the three methods. As shown in Fig. 19, it can be observed that due to the further flattening of the hill slope with *α*=1.2, the size of the recirculation zone is further reduced compared to the conditions of *α*=0.8 and 1.0. Despite this, the standard *k-ε* model still significantly underestimates the size of the recirculation zone, while the learned model demonstrates robust performance. As shown in Fig. 20, the skin friction predicted by the learned model closely aligns with the DNS results, significantly outperforming the predictions made by the *k-ε* model. Notably, both the learned model and DNS yield nearly identical reattachment point locations (*x/H*≈4.5), whereas the *k-ε* model predicts a reattachment location (*x/H* ≈3.7) that diverges considerably from the DNS results.

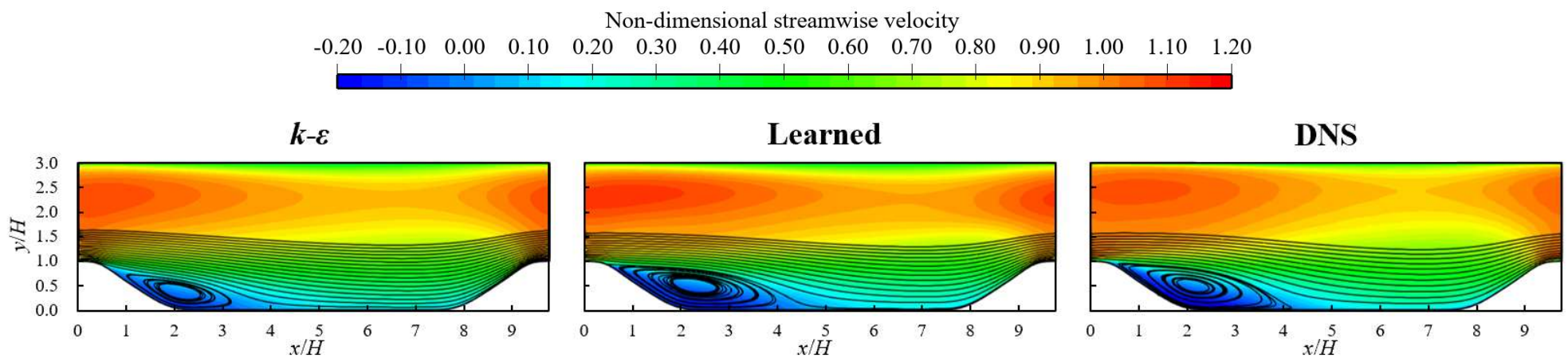

**Fig. 19** Non-dimensional streamwise velocity distribution and streamlines obtained by the standard *k-ε* model and the learned model, compared with DNS data (*α*=1.2, non-training case)

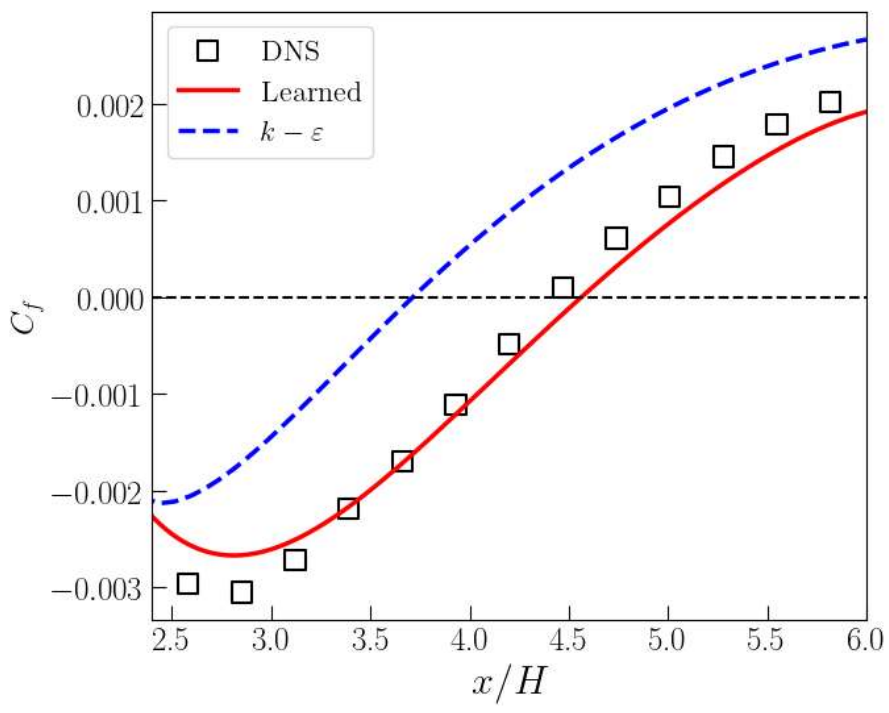

**Fig. 20** Skin friction distribution along the bottom wall obtained by the standard *k*-*ε* model and the learned model, compared with DNS data (*α*=1.2, training case)

Figure 21 presents the non-dimensional streamwise velocity profiles at different streamwise positions for the case *α*=1.2. The results indicate that the learned prediction model provides a more accurate representation of the streamwise velocity profiles compared to the standard *k*-*ε* model, particularly within the recirculation zone.

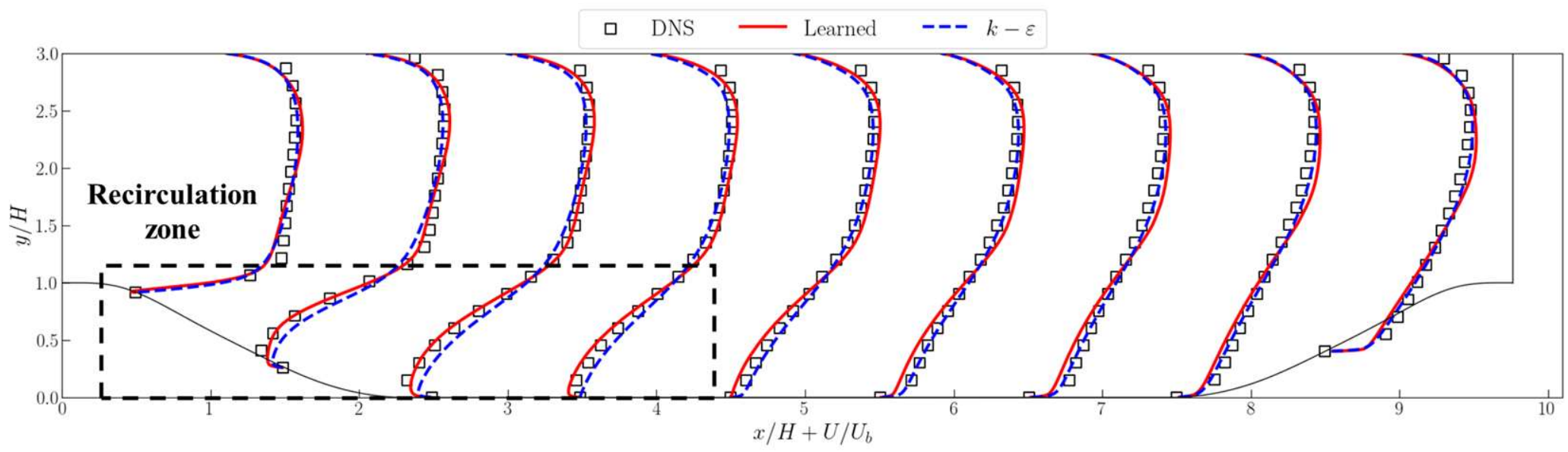

**Fig. 21** Non-dimensional streamwise velocity profile obtained by the standard *k*-*ε* model and the learned model, compared with DNS data (*α*=1.2, non-training case)

The results presented above qualitatively demonstrate the improvement in prediction accuracy of the learned model compared to the *k*-*ε* model for the flow over periodic hills. To quantitatively assess the learned model's effectiveness in capturing the velocity field, Table 4 provides the relative $l_2$-norm of error $E_i$ for the streamwise and vertical velocity fields obtained from the learned model and the *k*-*ε* model. These errors are calculated over the entire computational domain and within the recirculation zone separately, where the recirculation zone is approximately defined as *x/H*<5.0 and *y/H*<1.0. The relative $l_2$-norm of error is defined as [37]:

$$E_i = \frac{\left\| U_i - \hat{U}_i \right\|_2}{\left\| U_i \right\|_2} \tag{11}$$

where $\left\| \bullet \right\|_2$ represents the $l_2$-norm, $U_i$ and $\hat{U}_i$ denote the reference and predicted values of the velocity, respectively.

Table 5 demonstrates that under the three different $\alpha$, the learned model exhibits a reduction in prediction errors for both the streamwise velocity $U$ and the vertical velocity $V$ compared to the standard $k$-$\varepsilon$ model, particularly within the recirculation zone. For the streamwise velocity, the learned model's prediction error decreases from 2.03% to 0.89% across the entire computational domain and from 3.90% to 0.70% in the recirculation zone. This improvement is notable, considering that the $k$-$\varepsilon$ model maintains an error of around 9%. For the vertical velocity, the learned model shows a reduction in prediction error from 9.33% to 7.73% across the entire computational domain and from 16.16% to 6.78% in the recirculation zone, indicating a significant improvement compared to the $k$-$\varepsilon$ model. Furthermore, as $\alpha$ increases, the magnitude of the error reduction gradually diminishes. This trend can be attributed, on one hand, to the geometric deviation from the training conditions and, on the other hand, to the inherent reduction in prediction errors of the $k$-$\varepsilon$ model itself.

**Table 5** The relative $l_2$-norm of error (%) for three periodic hills cases

| | $\alpha$ | Full domain | | | Recirculation zone | | |
|---|---|---|---|---|---|---|---|
| | | $E_{\text{learned}}$ | $E_{k-\varepsilon}$ | $E_{k-\varepsilon}-E_{\text{learned}}$ | $E_{\text{learned}}$ | $E_{k-\varepsilon}$ | $E_{k-\varepsilon}-E_{\text{learned}}$ |
| $U$ | 0.8 | 8.17 | 10.20 | 2.03 | 6.69 | 10.59 | 3.90 |
| | 1.0 | 8.05 | 9.95 | 1.90 | 7.84 | 9.25 | 1.41 |
| | 1.2 | 8.60 | 9.49 | 0.89 | 7.48 | 8.18 | 0.70 |
| $V$ | 0.8 | 28.99 | 38.32 | 9.33 | 22.53 | 38.69 | 16.16 |
| | 1.0 | 25.03 | 33.66 | 8.63 | 14.79 | 30.27 | 15.48 |
| | 1.2 | 21.40 | 29.13 | 7.73 | 20.25 | 27.03 | 6.78 |

The previous analysis evaluated the performance of the learned model in the flow over periodic hills with three different steepness. The results indicate that the learned model provides more accurate predictions for Reynolds stress, velocity, and skin friction compared to the $k$-$\varepsilon$ model, and demonstrating a certain level of generalization capability. However, these two test cases primarily varying in geometry, which only partially reflects the learned model's generalization ability concerning boundary conditions. To further validate the model's generalization performance, the flow over a two-dimensional backward-facing step was investigated, which possesses an entirely different geometry and Reynolds number. This variation in Reynolds number will facilitate an assessment of the learned model's ability to generalize across diverse flow mechanisms.

The flow over a backward-facing step is a widely used benchmark for validating turbulence models in separated flows[38], as illustrated in Fig. 22. The flow enters from the upstream of the step, experiences separation after traversing a right-angle step, and subsequently reattaches downstream before exiting through the outlet. The height of the step is denoted as $H$. The non-dimensional length of the symmetry section is $L_{\text{is}}/H$ = 20. The non-dimensional distance from the end of the symmetry section to the step is $L_{iw}/H$ = 110, and the non-dimensional length from the step downstream to the outlet is $L_x/H$ = 50. The non-dimensional height of the computational domain is $L_y/H$ = 9. The Reynolds number based on the step height $H$ is $Re_H$ = 36000. The reference velocity $U_{ref}$ is defined as the velocity at the center channel near $x/H$=−4, which is used to compute the Reynolds number, non-dimensional velocity, and skin friction. The inlet velocity is specified using a fixed value boundary condition, while a zero-gradient boundary condition is applied at the outlet. To prevent potential incompatibility between the freestream flow and the wall boundary conditions, symmetry boundaries are applied at the top and bottom of the domain near the inlet, followed by no-slip wall conditions downstream. The numerical methods employed for this case are consistent with those described in Section 3, and the mesh independence has been verified.

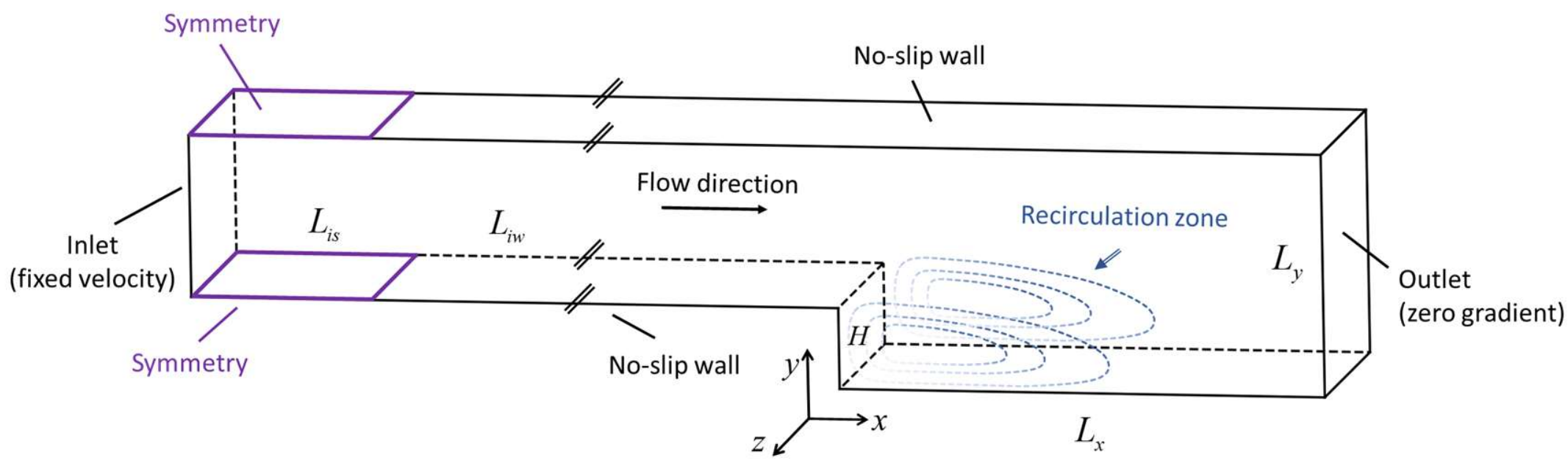

**Fig. 22** A sketch of the flow over a backward-facing step

Figure 23 presents the streamwise velocity distributions of the flow over a backward-facing step, as obtained from the learned model and the $k$-$\varepsilon$ model, at three distinct positions: the upstream attached zone ($x/H$=−4), the recirculation zone ($x/H$=1), and the downstream reattached zone ($x/H$=10). These results are compared with the experimental data from Driver and Seegmiller [39]. It can be observed that for both the upstream attached flow (left) and the downstream reattached flow (right), the streamwise velocity predicted by the learned model are similar to those of the $k$-$\varepsilon$ model, with both models underestimating the streamwise velocity in these regions. In these two areas, the learned model does not demonstrate a significant improvement in prediction accuracy compared to the $k$-$\varepsilon$ model. However, within the recirculation zone (middle), particularly in the near-wall area ($y/H$<1.0), the predictions of the learned model are significantly closer to the DNS results than those of the $k$-$\varepsilon$ model. However, as one moves away from the wall ($y/H$>1), the learned model's predictions become similar to those of the $k$-$\varepsilon$ model.

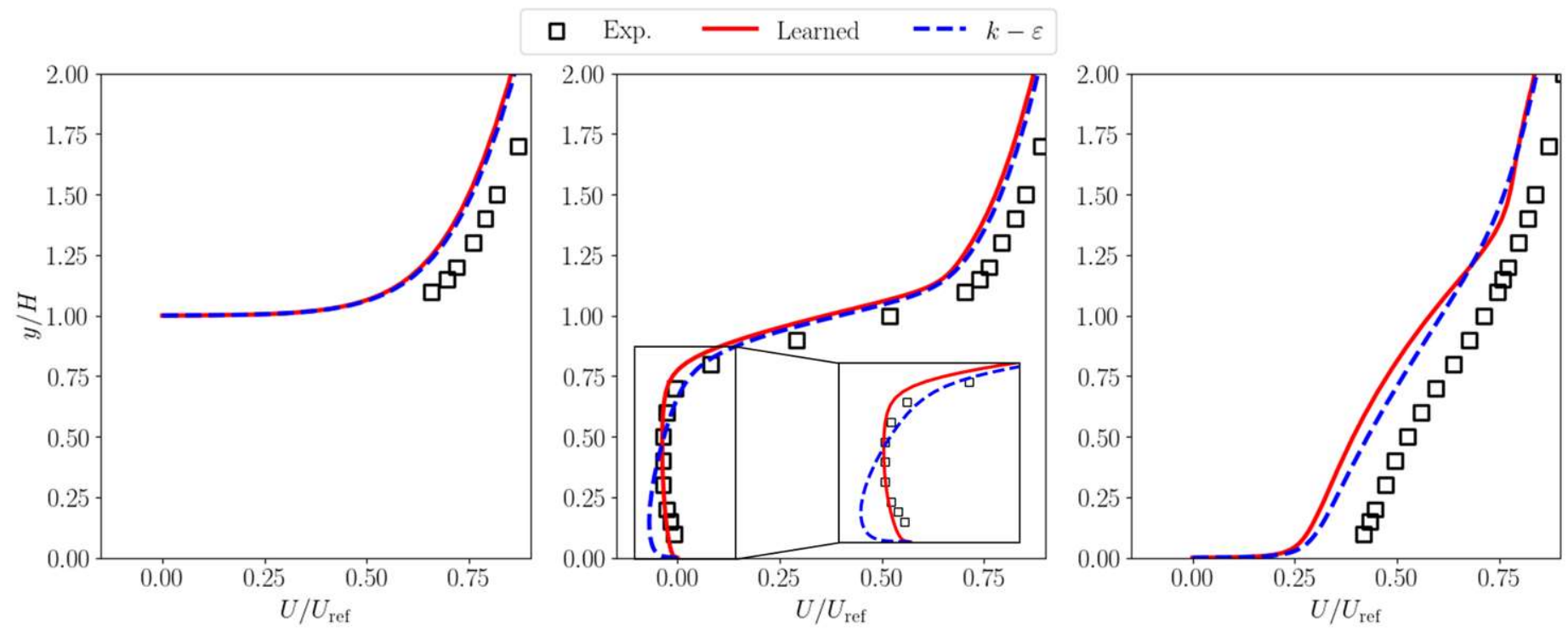

(a) $x/H$=-4 (b) $x/H$=1 (c) $x/H$=10

**Fig. 23** Non-dimensional streamwise velocity profile obtained by the standard *k-ε* model and the learned model at different streamwise positions, compared with experiment results (backward-facing step, non-training case)

Figure 24 presents the distribution of the skin friction across the range from *x/H*=0 to 12, as obtained from the learned model and the standard *k-ε* model, compared with the experimental results. It can be observed that the learned model provides a more accurate estimation of the skin friction, successfully predicting the peak value of $C_f$ in the recirculation zone and its streamwise location (*x/H*≈3.8). In contrast, the *k-ε* model predicts a larger peak value for $C_f$ and an earlier streamwise location (*x/H*≈3.0). Moreover, the DNS results indicate that the flow reattaches at approximately *x/H*≈6.0, while the learned model and the *k-ε* model predict reattachment at *x/H*≈5.5 and *x/H*≈5.0, respectively. These findings demonstrate that the learned model's predicted reattachment location aligns more closely with the experiment results.

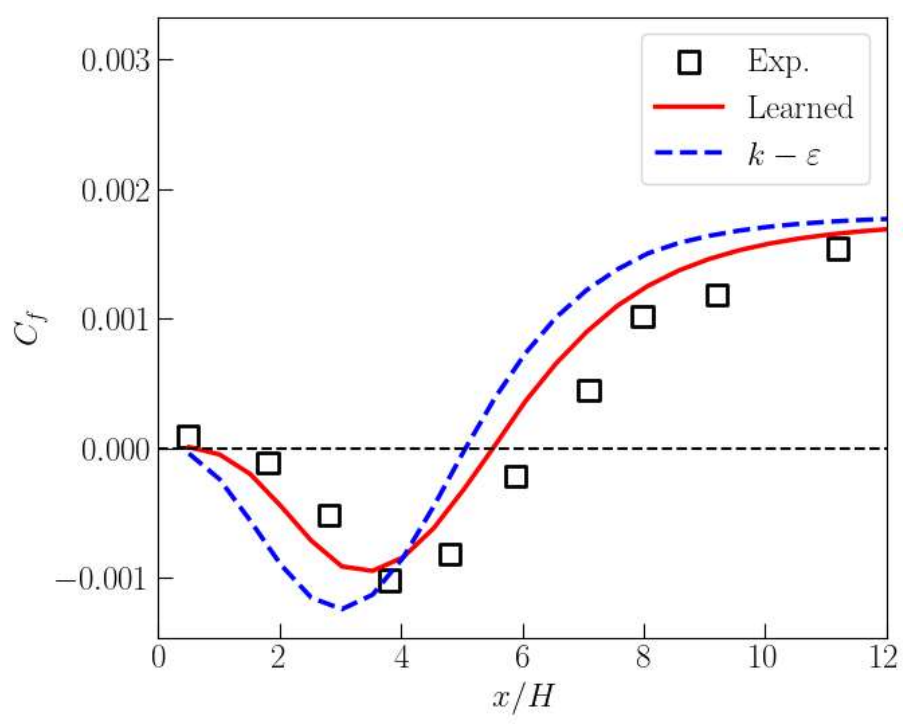

**Fig. 24** Skin friction distribution along the bottom wall obtained by the standard *k-ε* model and the learned model, compared with experiment result (backward-facing step, non-training case)

In summary, for the flow over a backward-facing step, which presents significant differences in geometry and Reynolds number compared to the training cases, the learned model still demonstrates improved prediction accuracy for both velocity and skin friction compared to the $k$-$\varepsilon$ model. This finding indicates that the learned model has considerable generalization capabilities.

## 5. Discussion

### 5.1. Effect of unit constraint

Unit constraints are central to the proposed framework, playing a key role in the symbolic regression learning process. Their effects on enhancing learning efficiency and improving the modeling capabilities for turbulent flow data, as well as the underlying mechanisms, warrant further in-depth discussion.

To quantitatively evaluate the influence of unit constraints on the algorithm's learning efficiency, an experiment without enforcing the unit constraints was conducted, using the same settings as in Section 3 for the Reynolds stress deviation learning. When the unit constraints are disabled, unit information is omitted from the observations in the RNN cell, and the output of the RNN cell is not checked for consistent dimensionality. Table 6 compares the CPU core time required to complete 200 training epochs with and without unit constraints. It is indicated that the version with unit constraints requires only about one-third of the time needed for the same number of epochs, demonstrating a significant improvement in algorithm runtime performance.

**Table 6** Runtime per core of the symbolic regression learning process for 200 epochs

| | Constrained | Unconstrained |
|---|---|---|
| Runtime per core | ~7.5min | ~20min |

Figure 25 illustrates the evolution over epochs of (a) the reward of the best expression in each batch and (b) the ratio of candidate expressions that are physically valid (i.e., those that satisfy the dimensional consistency principle) within each batch. The solid lines represent the results obtained with unit constraints imposed, whereas dashed lines correspond to the case where these constraints are disabled.

As shown in Fig. 25(a), disabling the unit constraints does not prevent the symbolic regression from achieving a higher reward—that is, lower error—in its learned expressions. In fact, the absence of this stringent constraint allows the algorithm to more rapidly generate expressions with lower error, which may give the false impression that the unconstrained algorithm has found superior expressions compared to the constrained version. However, upon examination of the units, the highest-reward expressions obtained without the constraints turn out to be dimensionally inconsistent and therefore physically invalid.

Figure 25(b) shows that when the unit constraints are disabled, only a very small fraction (less than 3%) of the candidate expressions generated in each batch satisfy the dimensional consistency principle. This indicates that the algorithm is merely achieving a better fit to the training data without capturing the underlying physics. In contrast, with unit constraints imposed, the ratio of physical expressions finally stabilizes at around 20–30%, demonstrating that these constraints effectively curb the generation of non-physical expressions. More importantly, as the number of epochs increases, the ratio of physical candidates rises rapidly before reaching a steady state, suggesting that the unit constraints actively guide the algorithm to produce expressions that obey the dimensional consistency principle. This behavior indirectly reflects that the neural network has, to some extent, learned the underlying physical principle related to dimensional consistency from the data.

Figure 25(b) further explains why the algorithm with unit constraints runs significantly faster than the unconstrained version. Specifically, when unit constraints are imposed, more than two-thirds of the expressions in each batch are zeroed out during the prior stage for failing the unit consistency check. This substantially reduces the computational cost in subsequent stages—such as reward calculation—resulting in a runtime reduction that matches the proportion of expressions discarded.

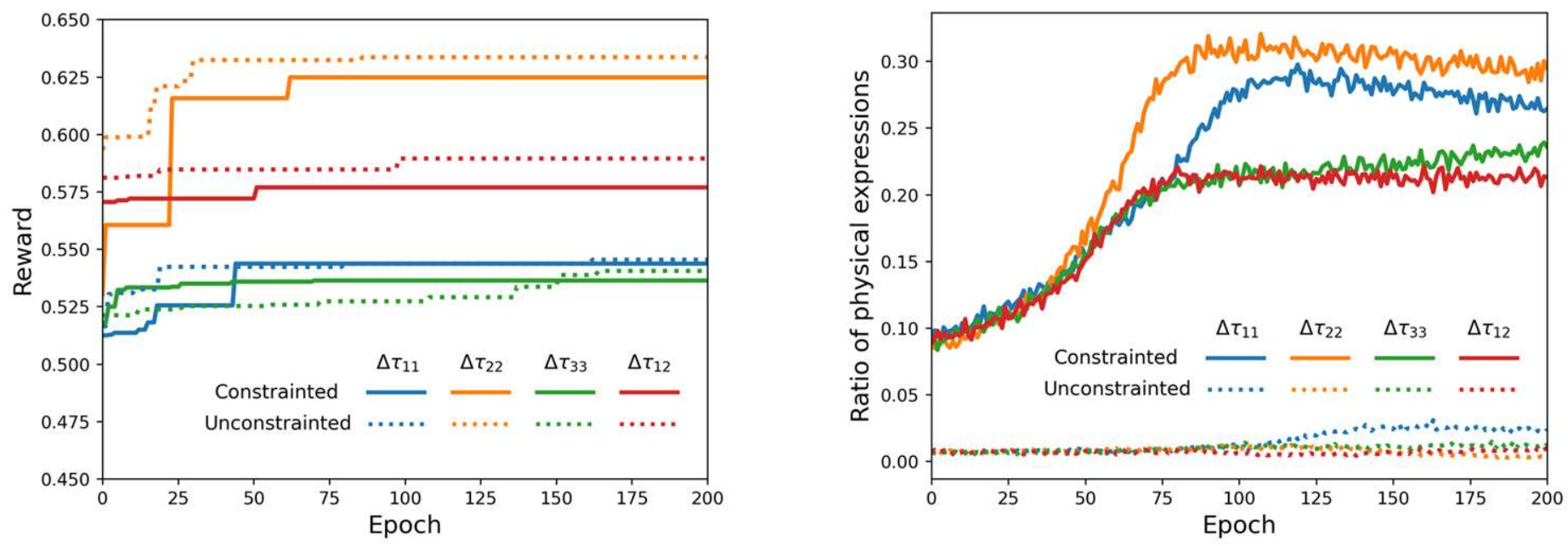

(a) best expression reward (b) ratio of physical expressions

**Fig. 25** Evolution of best expression reward and ratio of physical expressions over epochs

As an innovative attempt to apply unit constraints within symbolic regression for turbulence modeling, the results presented here demonstrate the potential of leveraging the principle of dimensional consistency to guide data-driven turbulence research in three key aspects.

First, the principle of dimensional consistency represents a fundamental rule governing the physical world, reflecting the inherent relationships among variables. Dimensional analysis based on this principle has long served as an essential tool for modeling unknown physical phenomena in fluid mechanics. By combining the powerful data-fitting capacity of machine learning with unit constraints, strong physical relationships are established among the flow quantities, equipping the symbolic regression with the ability to approximate physical laws from both the data and physics perspectives—an advantage beyond what purely data-driven approaches can offer.

Second, turbulence, as a typical high-dimensional complex dynamical problem, requires powerful modeling capabilities for accurate representation. Unit constraints enhance the search efficiency of symbolic regression, thereby making it feasible to utilize larger datasets, more input features, and more complex mathematical operators. This improvement will expand the potential of symbolic regression approaches in turbulence modeling research.

Finally, incorporating unit constraints ensures that the results of symbolic regression are physically valid in terms of dimensions. Compared to expressions obtained solely through data fitting, uncovering the physical significance encapsulated in these expressions is more meaningful, enhancing the possibility of extracting potentially valuable, unknown physical insights from the results.

It is important to emphasize that the inclusion of unit constraints is only the beginning; the algorithmic framework presented here also allows for the integration of any additional physical prior constraints, offering promising prospects for further enhancing the capabilities of symbolic regression in turbulence modeling research in the future.

### 5.2. Physical interpretation of the learning result

The primary advantage of symbolic regression lies in its interpretability, which, in turbulence modeling, helps researchers uncover the underlying physical mechanisms, and provide concrete guidance for more refined and accurate models. The following analysis discusses the impact mechanism and physical significance of the Reynolds stress deviation correction formulas obtained in this study.

Based on the flow results shown in Figures 10, 14, and 18, it is evident that the regions above the recirculation zone—the free shear layer—and the windward hillside are where the LEVM exhibits its primary errors and where the Reynolds stress corrections play a dominant role. In the free shear layer, the high shear strain generates significant stress production and leads to elevated streamwise normal stress $\tau_{11}$, which the linear eddy viscosity model underestimates. Since the streamwise component dominates, the production of turbulent kinetic energy shows a distribution similar to that of the streamwise stress. The correction $\tau_{11}$, by incorporating an additional $k$-driven term ($q_2$), alleviates the underestimation of the streamwise stress in free shear flows. A similar corrective mechanism applies to the streamwise Reynolds stress on the windward hillside.

In the case of $\tau_{22}$, symbolic regression yielded an expression nearly opposite to that for $\Delta\tau_{11}$. This may reflect an energy conservation principle or simply indicate that a negative turbulent kinetic energy distribution better matches the observed error distribution.

For the normal stress $\tau_{33}$ previous studies (e.g., Fröhlich et al.[40]) have shown that near the wall on the windward hillside, energy is redistributed from the streamwise stress $\tau_{11}$ and wall-normal stress $\tau_{22}$ to the spanwise stress, leading to a notable increase in $\tau_{33}$. This mechanism, driven by pressure–strain correlation, is not captured by the LEVM, resulting in an underestimation of the spanwise stress. As shown in Fig. 8, the correction $\Delta\tau_{33}$ successfully provides a positive adjustment to the spanwise stress, and the automatically identified switching function confines this correction to the near-wall region. Remarkably, symbolic regression not only applies the correct positive correction to the spanwise stress but also, through the $q_3$ quadratic term in $\Delta\tau_{11}$ and the denominator in $\Delta\tau_{22}$, reduces the values of $\Delta\tau_{11}$ and $\Delta\tau_{22}$ near the wall, thereby capturing the underlying energy redistribution mechanism.

For the shear stress $\Delta\tau_{12}$, figures 10, 14, and 18 indicate that the $k$–$\varepsilon$ model exhibits anomalous predictions primarily on the windward hillside. The correction $\Delta\tau_{12}$ effectively ameliorates this anomaly. Considering the expression structure and the information in Fig. 8, the turbulent kinetic energy convection term $q_8$ shows a distribution on the windward hillside that closely aligns with that of $\Delta\tau_{12}$. This suggests that $q_8$ plays a dominant role in this correction and that the anomalous behavior of $\Delta\tau_{12}$ could be related to turbulent kinetic energy convection.

Nonetheless, the current model has three main limitations. First, to balance the high-error region on the windward hillside, the global learning strategy over the entire flow field resulted in an excessive correction of $\tau_{22}$ in the free shear layer, leading to an underestimation of $\tau_{22}$; future work should consider region-specific modeling to address this issue. Second, the current predictive model does not sufficiently correct the spanwise normal stress $\tau_{33}$, indicating that the introduction of a more refined switching function may be necessary.

## 6. Conclusion

In this study, we proposed a novel unit-constrained turbulence modeling framework using symbolic regression to improve the prediction accuracy of LEVMs for large separated turbulence. The proposed framework employs symbolic regression to learn explicit equations between Reynolds stress discrepancies and mean flow features from RANS baseline data and high-fidelity DNS turbulence data. By incorporating the obtained equations as correction terms into the constitutive equations, a new learned turbulence model is obtained. Unit consistency constraints are applied during the symbolic regression to enhance physical realizability.

To validate the effectiveness of the proposed turbulence modeling framework, the flow over periodic hills is used as the training case. Testing cases included flows over periodic hills with varying steepness and a backward-facing step featuring different geometries and Reynolds numbers. The findings reveal that the proposed framework exhibits superior modeling performance for large separated turbulence compared to the standard *k*-*ε* model. In both training and non-training cases with different geometries, the learned model consistently predicts more realistic Reynolds stress distributions, more accurate velocity fields and skin friction than the standard *k*-*ε* model. Notably, even in the flow over backward-facing step, which presents significant differences in geometry and Reynolds number from the training cases, the learned model exhibits noteworthy performance. This indicates that the learned model not only effectively captures the influence of boundary conditions on the flow but also successfully extracts the underlying flow mechanisms present in the data, thereby demonstrating a considerable generalization capability.

Some further in-depth discussions were conducted. First, the intrinsic effects of unit constraints on the efficiency and learning process of symbolic regression are analyzed. The results indicate that the algorithm's runtime under unit constraints is reduced to approximately one-third of that in the unconstrained case, and these constraints will guide the symbolic regression algorithm to actively learn expressions that adhere to dimensional homogeneity. Next, we examined the physical implications of the Reynolds stress discrepancy expressions derived from the model. One surprising finding is that the model may have successfully captured the energy redistribution mechanism on the windward hillside. In future studies, more refined subregional modeling and the design of appropriate switching functions may be key to further enhancing the model's performance.

This study has three main limitations. First, compared to the *k*–*ε* model, the learned model shows marginal benefits in some cases; for instance, in the back-ward facing step case, the model appreciably reduces prediction errors only for the velocity distribution within the recirculation region. This limitation may stem from the learned model focusing solely on flow-separation patterns, and incorporating training data covering a broader range of flow types could help improve its performance. Second, since the mean flow features used to construct the corrective expression are computed from local flow quantities, the non-local characteristics of the flow are not accounted for, and thus the non-local transport mechanisms of turbulence might not be directly reflected in the model. Finally, this study primarily addresses statistical two-dimensional separated turbulent flows, which represent a relatively simplified flow scenario compared to those encountered in practical engineering applications. More complex flow scenarios, such as shock-boundary layer interactions, may require the construction of additional and more complex mean flow features to capture the underlying intricate mechanisms.

Current research represents an initial step toward developing a more general turbulence model via symbolic regression. Future efforts will aim to progressively enhance the model's robustness and extend its applicability to a broader range by improving on three key aspects. First, the training dataset will be broadened to include a wider variety of flow types, thereby further the enhancing generalizability of the learned model. Second, more physical constraints, such as conservation laws, will be introduced to improve physical modeling capability. Finally, differential operators will be incorporated into the symbolic regression framework, thereby endowing the framework with the capability to learn partial differential equations and establishing a closure model that accounts for non-local transport effects.

## Funding Information

We confirm that there has been no significant financial support for this work that could have influenced its outcome.

## Conflict of Interest

The authors declare that they have no conflict of interest.

## Data Availability

The datasets generated during and/or analyzed during the current study are available from the corresponding author on reasonable request.

## Appendix A: Geometry of the parametric periodic hills

The geometry of the first hill upstream in the periodic hills ($\alpha$=1.0) is given by the following equation [35]:

$$\begin{cases} \hat{y} = \min\left(1; 1 + 2.42\times10^{-4}\hat{x}^2 - 7.588\times10^{-5}\hat{x}^3\right) & ,\hat{x} \in [0, 0.3214] \\ \hat{y} = 0.8955 + 3.484\times10^{-2}\hat{x} - 3.629\times10^{-3}\hat{x}^2 + 6.749\times10^{-5}\hat{x}^3 & ,\hat{x} \in (0.3214, 0.5] \\ \hat{y} = 0.9213 + 2.931\times10^{-2}\hat{x} - 3.234\times10^{-3}\hat{x}^2 + 5.809\times10^{-5}\hat{x}^3 & ,\hat{x} \in (0.5, 0.7143] \\ \hat{y} = 1.445 - 4.927\times10^{-2}\hat{x} + 6.95\times10^{-4}\hat{x}^2 - 7.394\times10^{-6}\hat{x}^3 & ,\hat{x} \in (0.7143, 1.071] \\ \hat{y} = 0.6401 + 3.123\times10^{-2}\hat{x} - 1.988\times10^{-3}\hat{x}^2 + 2.242\times10^{-5}\hat{x}^3 & ,\hat{x} \in (1.071, 1.429] \\ \hat{y} = \max\left(0; 2.0139 - 7.18\times10^{-2}\hat{x} + 5.875\times10^{-4}\hat{x}^2 + 9.553\times10^{-7}\hat{x}^3\right) & ,\hat{x} \in (1.429, 1.929] \end{cases} \tag{A1}$$

where $\hat{x}$ and $\hat{y}$ represent the non-dimensional coordinates of the hill, with the expressions given by $\hat{x} = x / H$ and $\hat{y} = y / H$, respectively. The geometry of the second hill downstream is derived by mirroring the first hill along the *y*-axis and then translating it along the *x*-axis.

The relationship between the total length $L_x$ of the domain and $\alpha$ is as follows:

$$L_x / H = 3.858\alpha + 5.142 \tag{A2}$$

As $\alpha$ changes, $L_x$ also changes accordingly, and the shape of the hills and the bottom wall will be stretched along the *x*-axis to accommodate the variation in $L_x$.